\documentclass[10pt,journal,letterpaper,twoside,twocolumn,nofonttune]{IEEEtran}

\usepackage{tabularx}

\usepackage{amsmath,amsfonts}
\usepackage{algorithmic}
\usepackage{algorithm}
\usepackage{array}
\usepackage{float}
\usepackage{multirow}
\usepackage{textcomp}
\usepackage{stfloats}
\usepackage{url}
\usepackage{verbatim}
\usepackage{graphicx}
\usepackage{cite}
\usepackage{amsmath, xparse}
\usepackage[utf8]{inputenc}
\usepackage{xcolor}
\usepackage{cite}
\usepackage{amsmath,amssymb,amsfonts}
\usepackage{algorithmic}
\usepackage{graphicx}
\usepackage{textcomp}
\usepackage{bm}
\usepackage{float}
\usepackage{xcolor}
\usepackage{soul}
\usepackage{xfrac}
\usepackage{subfigure}

\usepackage{soul}

\definecolor{LightGray}{gray}{0.8}
\definecolor{Orange}{rgb}{1.0, 0.31, 0.0}
\definecolor{Green}{rgb}{0.3, 1.0, 0.3}
\definecolor{Blue}{rgb}{0.75,0.75,1}
\definecolor{Magenta}{rgb}{255, 0, 255}
\definecolor{Cyan}{rgb} {0, 255, 255}
\newcommand{\hlg}[2][LightGray]{{\sethlcolor{#1}\hl{#2}}}

\newcommand{\bea}{\begin{eqnarray}}
\newcommand{\beal}[1]{\begin{eqnarray}\label{#1}}
\newcommand{\eea}{\end{eqnarray}}
\def\balg#1#2\ealg{\begin{align}\label{#1}#2\end{align}}
\def\balgnl#1\ealgnl{\begin{align*}#1\end{align*}}

\newcommand{\E}{{\mathbf E}}

\ifCLASSINFOpdf
\else
\fi
\begin{document}
%
\title{A Time-Domain Method of Auxiliary Sources for Analyzing Transient Electromagnetic Interactions with GSTC-Modeled Metasurfaces}

%

\author{Minas~Kouroublakis, Nikolaos~L.~Tsitsas, \IEEEmembership{Senior Member, IEEE}, and Yehuda Leviatan, \IEEEmembership{Life Fellow, IEEE} 
\thanks{\emph{Corresponding author: Nikolaos~L.~Tsitsas}}
\thanks{Minas~Kouroublakis and Nikolaos~L.~Tsitsas are with the School of Informatics, Aristotle University of Thessaloniki, 54124 Thessaloniki, Greece
(e-mails: mkour2000@yahoo.com; ntsitsas@csd.auth.gr).
}
\thanks{Yehuda Leviatan is with Department of Electrical and Computer Engineering, Technion-Israel Institute of Technology, Haifa, 32000, Israel (e-mail: leviatan@technion.ac.il)}}

%
%

\markboth{IEEE TRANSACTIONS ON ANTENNAS AND  PROPAGATION}%
{IEEE TRANSACTIONS ON ANTENNAS AND  PROPAGATION}


%


\maketitle

\begin{abstract}
This paper presents a time-domain (TD) formulation for modeling the transient electromagnetic response of two-dimensional (2-D) metasurfaces using the Method of Auxiliary Sources (MAS) combined with the Generalized Sheet Transition Condition (GSTC).  In the proposed approach, the frequency-domain impedance-type GSTC is transformed into a causal, convolution-based TD representation and integrated within the MAS formulation. The method supports both $\mathrm{TM}_z$ and $\mathrm{TE}_z$ polarizations and accommodates general transient excitations, including pulsed plane waves and localized sources. Numerical experiments are presented for several representative cases—anisotropic graphene, black phosphorus, and an artificial Lorentzian metasurface—demonstrating the method's accuracy, stability, and versatility. Validations against both a frequency-domain MAS solver and full-wave simulations performed in 
commercial software confirm the correctness of the results, establishing the proposed TD-MAS GSTC formulation as an efficient meshless tool for broadband TD analysis of metasurfaces.
\end{abstract}
%


\begin{IEEEkeywords}
Time-domain electromagnetics, metasurfaces, generalized sheet transition conditions (GSTC), method of auxiliary sources (MAS), anisotropic graphene, black phosphorus, Lorentzian metasurfaces, transient analysis.

\end{IEEEkeywords}
%

%
\IEEEpeerreviewmaketitle

\section{Introduction}
\label{sec:Introduction}
\IEEEPARstart{T}{ime}-domain (TD) electromagnetics has become increasingly important owing to a broad range of applications in radar systems~\cite{luo2021timedomain, muppala2024fast}, geophysical exploration~\cite{carrasco2022time,  parshin2021lightweight}, non-destructive testing~\cite{ guo2022review}, and biomedical imaging~\cite{wang2021overview}.
In recent years, it has also emerged as a powerful tool for investigating electromagnetic (EM) metasurfaces—two-dimensional (2-D) arrays of subwavelength features designed to control the behavior of EM waves \cite{holloway2012overview, achouri2021electromagnetic}. 
TD modeling of such structures is of growing relevance, as metasurfaces need to be designed for broadband, and transient wave manipulation in applications like dynamic beam steering \cite{nadi2024beam}, pulse shaping \cite{geromel2023compact}, and temporal modulation \cite{wang2022pseudorandom,mikheeva2022space}.
TD techniques are advantageous in this context, as they rely on short pulses exciting a wide frequency range, enabling the broadband response of a metasurface to be captured in a single simulation or experiment. The pulse duration governs the spectral bandwidth, with shorter pulses yielding broader frequency coverage; this is essential for analyzing metasurfaces under realistic, time-varying excitations.

A broad range of numerical techniques have been proposed for solving TD EM problems~\cite{ren2022advances}, encompassing applications from transient scattering to the modeling of dispersive and anisotropic materials. Among the most established approaches, the Finite-Difference Time-Domain (FDTD) method remains a popular choice owing to its straightforward formulation and ability to capture broadband dynamics in complex geometries~\cite{gedney2022introduction, teixeira2023finite}.
The Time-Domain Finite Element Method (TD-FEM) represents another mature option and is widely available in commercial solvers such as COMSOL Multiphysics~\cite{Comsol}.  
Additional alternatives include Galerkin-based TD formulations~\cite{alvarez2012spurious, angulo2015discontinuous}, as well as TD surface- and volume-integral-equation methods~\cite{takahashi2023fast, chen2020explicit}, all of which have been extensively developed for transient EM analysis. However, these schemes typically require volumetric or surface meshing, which can lead to significant computational overhead—particularly when modeling electrically thin, spatially dispersive layers such as metasurfaces.
The Method of Auxiliary Sources (MAS)~\cite{papakanellos2024method,kaklamani2002}, 
also referred to as the Method of Fundamental Solutions (MFS)~\cite{Fairweather2003, Karageorghis2025} or the Source Model Technique (SMT)~\cite{leviatan1988generalized, tsitsas2018}, offers a meshless alternative circumventing this limitation. It has demonstrated adequate accuracy and efficiency in TD modeling of 2-D EM structures~\cite{ludwig2006towards, ludwig2007source, ludwig2008time, ludwig2011source}. More recently, a TD--MAS implementation was developed incorporating the Standard Impedance Boundary Condition (SIBC) in~\cite{TDSIBCarxiv}, and the Impedance Matrix Boundary Condition (IMBC) in \cite{TD_shielding}, providing foundations for analyzing more general metasurfaces.

The \emph{Generalized Sheet Transition Condition} (GSTC) gives a rigorous theoretical framework for modeling metasurfaces by relating the discontinuities of the tangential EM fields across an infinitesimally thin sheet to its effective electric and magnetic surface susceptibilities \cite{achouri2015general}. With this approach, a physically thin but volumetric metamaterial layer is replaced by an equivalent zero-thickness sheet, reducing modeling complexity and avoiding the dense meshing requirements typically associated with volumetric representations. The GSTC framework has been extensively implemented in the frequency domain using several numerical schemes, including the Finite Element Method (FEM)~\cite{sandeep2017finite}, integral equation methods ~\cite{dugan2022accelerated, smy2021ie}, and the MAS ~\cite{wang2020simulation}, for planar and cylindrical geometries. In parallel, TD GSTC-based techniques have also been developed—most notably using the FDTD method~\cite{vahabzadeh2017generalized,ha2025fdtd,smy2020fdtd} and Galerkin TD methods \cite{tian2022modeling}.
Despite these advances, most existing TD implementations remain grid-based, motivating the development of meshless formulations for the accurate and efficient modeling of metasurfaces under transient excitations.

In this work, we extend the TD-MAS to incorporate the impedance-type GSTC \cite{wang2020simulation}, leading to a fully meshless and broadband formulation for transient analysis of metasurfaces. The proposed TD MAS--GSTC method directly models the field discontinuities imposed by electric and magnetic surface susceptibilities, enabling accurate simulation of metasurface behavior without volumetric discretization or numerical meshing.
To assess its performance, we investigate three representative scenarios: (i) a magnetostatically biased anisotropic graphene sheet \cite{kouroublakis2023shielding, jang2023unified}, (ii) a black phosphorus layer exhibiting in-plane anisotropy \cite{low2014plasmons,jang2022efficient}, and (iii) an artificial Lorentzian metasurface characterized by dispersive electric and magnetic susceptibilities \cite{smy2020fdtd}.
These examples collectively validate the accuracy, stability, and generality of the proposed formulation while providing implementation guidelines for practical metasurface modeling in the TD.


The proposed formulation is developed under a set of assumptions that determine the range of validity of the method. The metasurface is assumed to be infinite along the tangential ($y$ and $z$) directions and to have zero thickness along the normal ($x$) direction. Under this assumption, the metasurface is represented only through field discontinuities across its plane and the problem is therefore treated as two-dimensional. Hence, only tangential field components and equivalent surface currents are considered, while normal polarizations, finite-size effects, and thickness-related phenomena are not included. Furthermore, the metasurface response is assumed to be local, meaning that the induced surface polarization at each point depends only on the local tangential fields. Spatial dispersion is therefore neglected. Temporal dispersion is included through the convolution relations, whereas space--time modulation of the metasurface parameters is outside the scope of the present work.

The remainder of this paper is organized as follows.  
Section~\ref{sec:theory} presents the theoretical framework of the proposed TD MAS--GSTC formulation, including the derivation of the impedance-type GSTC equations, their frequency-to-time domain transformation, and integration within the MAS scheme.  
Section~\ref{sec:Numerical} reports a comprehensive set of numerical experiments, demonstrating the method’s accuracy and stability across three representative configurations: magnetically biased graphene, black phosphorus, and an artificial Lorentzian metasurface. Finally, Section~\ref{sec:conclusions} summarizes the main outcomes of this study and discusses prospective research directions for extending the MAS--GSTC approach to more general metasurface geometries and three-dimensional configurations.

\section{The TD-MAS for Zero Thickness Metasuraces}
\label{sec:theory}

This section presents the theoretical foundation for applying the MAS in conjunction with the impedance-type GSTCs for modeling transient EM interactions with zero-thickness metasurfaces. The formulation begins with the frequency-domain (FD) representation of the impedance-type GSTC, in which the tangential electric fields on both sides of the sheet are related to the tangential magnetic fields by a surface susceptibility matrix. Then, the transformation of the GSTC into the TD is addressed via the inverse Fourier transform (IFT), properly accounting for the convolutional behavior of the TD operators. Finally, the integration of the TD impedance-type GSTC within the MAS framework is presented, enabling accurate and efficient simulation of broadband transient scattering phenomena from general anisotropic or bianisotropic metasurfaces.

\subsection{Frequency-Domain Analysis of GSTC}
\begin{figure}[htb!]
\centering
\subfigure[]{\includegraphics[width=.25\textwidth]{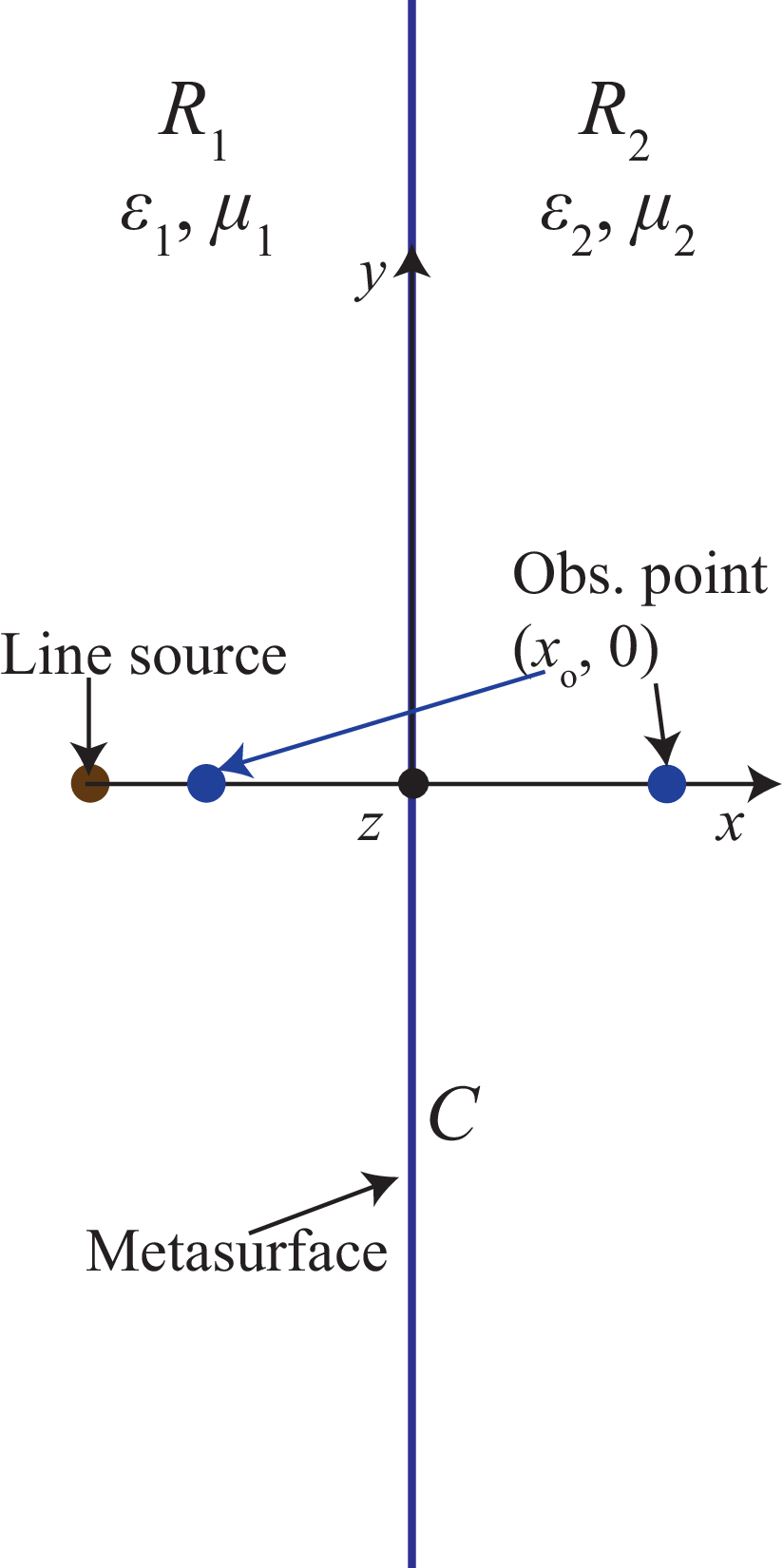}} 
\hfill
    \subfigure[]{\includegraphics[width=0.22\textwidth]{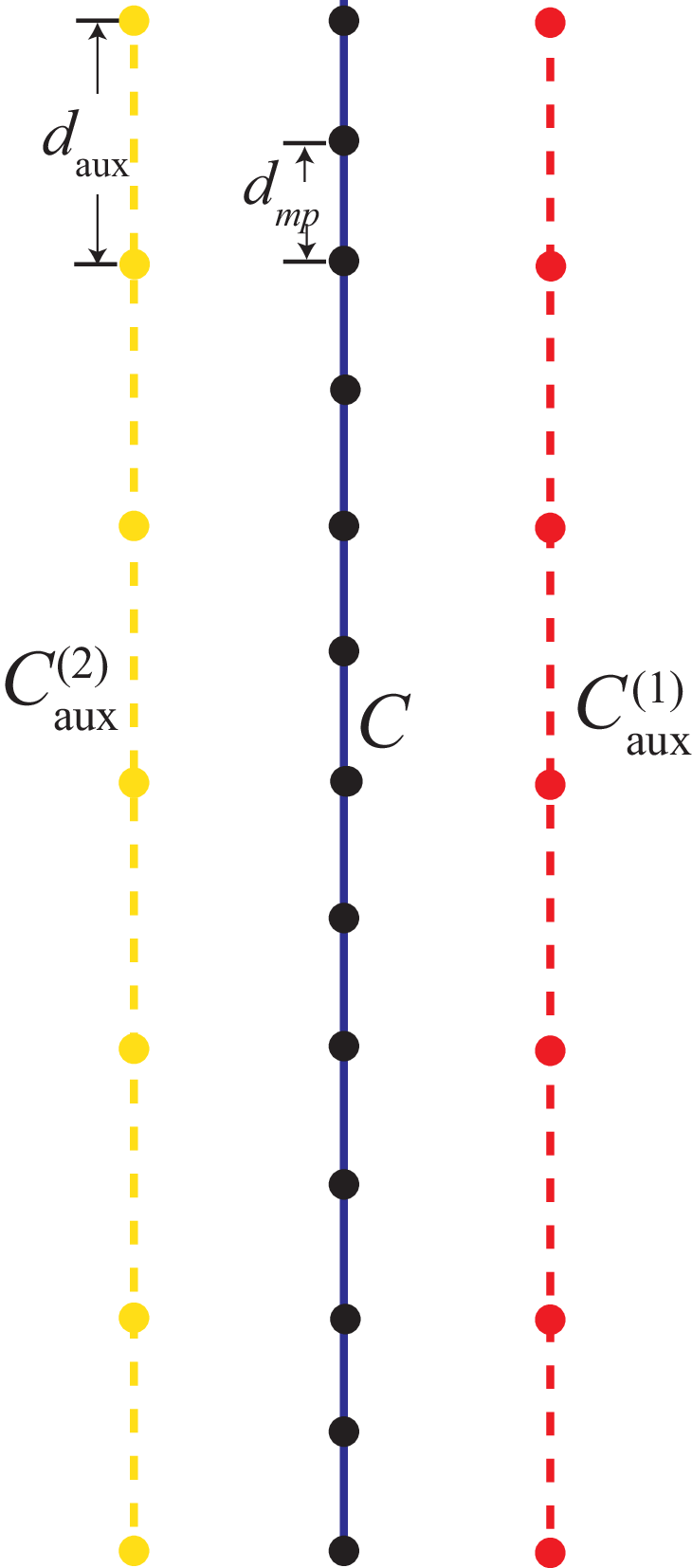}}
    \hfill  
\caption{(a) 2-D metasurface and (b) application of TD MAS GSTC. Black dots represent matching points on $C$, red (yellow) dots sources on  $C^{(1)}_{\rm aux}$ $(C^{(2)}_{\rm aux})$.}
\label{fig:geometry_MAS}\unskip
\end{figure}

Figure~\ref{fig:geometry_MAS}(a) illustrates a zero-thickness infinite planar metasurface lying on the \( yz \)-plane. The surrounding space is divided into two homogeneous regions \( R_1 \) and \( R_2 \) of permittivity \( \varepsilon_0 \) and permeability \( \mu_0 \). An incident field originating from region \( R_1 \) impinges on the metasurface, corresponding either to a plane wave or to a cylindrical wave radiated by an infinitely-long electric current filament near the sheet. When illuminated, the metasurface imposes specified discontinuities on the tangential fields at its location, thereby controlling the transmitted and reflected wavefronts according to its surface parameters. 

The volumetric response of a metasurface can be homogenized and represented by effective surface parameters, which are expressed by susceptibilities relating the induced polarization densities to the average tangential fields at the interface. The behavior of the metasurface---including reflection, transmission, and polarization conversion---is characterized by its susceptibility tensors. The impedance-type GSTC \cite{wang2020simulation} relates the tangential electric field components on both sides of the metasurface to the tangential magnetic field ones as follows:
\begin{equation}
[L]
\begin{bmatrix}
E_{1y} \\[3pt]
E_{2y} \\[3pt]
E_{1z} \\[3pt]
E_{2z}
\end{bmatrix}
=
[R]
\begin{bmatrix}
H_{1z} \\[3pt]
H_{2z} \\[3pt]
H_{1y} \\[3pt]
H_{2y}
\end{bmatrix},
\label{eq:GSTC_matrix}
\end{equation}
with
\begin{align}
[L] &=
\begin{bmatrix}
A_{11} & A_{11} & A_{12} & A_{12} \\
A_{21} & A_{21} & A_{22} & A_{22} \\
D_{11} & D_{11} & D_{12}+1 & D_{12}-1 \\
D_{21}-1 & D_{21}+1 & D_{22} & D_{22}
\end{bmatrix},
\label{eq:Lmatrix} \\[6pt]
[R] &= -
\begin{bmatrix}
B_{12}-1 & B_{12}+1 & B_{11} & B_{11} \\
B_{22} & B_{22} & B_{21}+1 & B_{21}-1 \\
C_{12} & C_{12} & C_{11} & C_{11} \\
C_{22} & C_{22} & C_{21} & C_{21}
\end{bmatrix}
\label{eq:Rmatrix}
\end{align}
\noindent where $A_{ij}$, $B_{ij}$, $C_{ij}$, and $D_{ij}$ are the elements of the dyadics

\begin{align}
[A] &= \tfrac{1}{2}\mathrm{i}\omega\varepsilon_0
\begin{bmatrix}
\chi^{yy}_{ee} & \chi^{yz}_{ee} \\
\chi^{zy}_{ee} & \chi^{zz}_{ee}
\end{bmatrix}, &
[B] &= \tfrac{1}{2}\mathrm{i}k_0
\begin{bmatrix}
\chi^{yy}_{em} & \chi^{yz}_{em} \\
\chi^{zy}_{em} & \chi^{zz}_{em}
\end{bmatrix}, \\[6pt]
[C] &= \tfrac{1}{2}\mathrm{i}\omega\mu_0
\begin{bmatrix}
\chi^{yy}_{mm} & \chi^{yz}_{mm} \\
\chi^{zy}_{mm} & \chi^{zz}_{mm}
\end{bmatrix}, &
[D] &= \tfrac{1}{2}\mathrm{i}k_0
\begin{bmatrix}
\chi^{yy}_{me} & \chi^{yz}_{me} \\
\chi^{zy}_{me} & \chi^{zz}_{me}
\end{bmatrix}.
\end{align}
with $\chi_{ee}$, $\chi_{mm}$, $\chi_{em}$, and $\chi_{me}$ the electric and magnetic surface susceptibilities describing the metasurface response to electric or magnetic excitations, respectively, while $k_0$ denotes the free-space wavenumber. We note that normal polarization densities are assumed to be negligible, given the subwavelength thickness of the metasurface.
Consequently, the four susceptibility tensors $[A]$, $[B]$, $[C]$, and $[D]$ each exhibit a $2\times 2$ form, representing purely tangential field interactions. The subscripts “1” and “2” on the fields refer to regions $R_1$ (for incident and reflected fields) and $R_2$ (for transmitted fields), respectively.

\subsection{Time-Domain Transformation of the GSTC}

The transformation of the impedance-type GSTC from the FD to the TD requires the application of the inverse Fourier transform (IFT) to Eq.~(\ref{eq:GSTC_matrix}). 
It is important to recall that the product of two functions in the frequency domain corresponds to the convolution of their respective inverse Fourier transforms in the time domain. 
Hence, each element of the surface susceptibility dyadics—such as $\chi_{ee}$, $\chi_{mm}$, $\chi_{em}$, and $\chi_{me}$—which multiply the field quantities in the FD formulation, becomes a TD convolution operator acting on the corresponding field components. 
These convolutional terms reflect the dispersive nature of the metasurface, ensuring that the GSTC accurately captures both instantaneous and delayed EM responses within the TD analysis. 
By applying the IFT to (\ref{eq:GSTC_matrix}), the FD impedance-type GSTC is converted into its TD counterpart, leading to the system of convolution-type integral equations
\begin{subequations}
\label{eq:integral_equations}
\begin{multline}
a_{11}(t)\ast E_{1y}(t)+a_{11}(t)\ast E_{2y}(t)+\\a_{12}(t)\ast E_{1z}(t)+a_{12}(t)\ast E_{2z}(t)=\\
-\Bigl (b_{12}(t)\ast H_{1z}(t)-H_{1z}(t)+b_{12}(t)\ast H_{2z}(t)\\+H_{2z}(t)+b_{11}(t)\ast H_{1y}(t)+b_{11}(t)\ast H_{2y}(t)\Bigr),
\end{multline}
\begin{multline}
a_{21}(t)\ast E_{1y}(t)+a_{21}(t)\ast E_{2y}(t)+\\a_{22}(t)\ast E_{1z}(t)+a_{22}(t)\ast E_{2z}(t)=\\
-\Bigl(b_{22}(t)\ast H_{1z}(t)+b_{22}(t)\ast H_{2z}(t)+\\b_{21}(t)\ast H_{1y}(t)+H_{1y}(t)+b_{21}(t)\ast H_{2y}(t)-H_{2y}(t)\Bigr),
\end{multline}
\begin{multline}
d_{11}(t)\ast E_{1y}(t)+d_{11}(t)\ast E_{2y}(t)+\\d_{12}(t)\ast E_{1z}(t)+E_{1z}(t)+d_{12}(t)\ast E_{2z}(t)-E_{2z}(t)=\\
-\Bigl(c_{12}(t)\ast H_{1z}(t)+c_{12}(t)\ast H_{2z}(t)+\\c_{11}(t)\ast H_{1y}(t)+c_{11}(t)\ast H_{2y}(t)\Bigr),
\end{multline}
\begin{multline}
d_{21}(t)\ast E_{1y}(t)-E_{1y}(t)+d_{21}(t)\ast E_{2y}(t)+\\E_{2y}(t)+d_{22}(t)\ast E_{1z}(t)+d_{22}(t)\ast E_{2z}(t)=\\
-\Bigl(c_{22}(t)\ast H_{1z}(t)+c_{22}(t)\ast H_{2z}(t)+\\c_{21}(t)\ast H_{1y}(t)+c_{21}(t)\ast H_{2y}(t)\Bigr).
\end{multline}
\end{subequations}
\noindent Here, the functions $a_{ij}(t)$, $b_{ij}(t)$, $c_{ij}(t)$, and $d_{ij}(t)$ represent the IFTs of the corresponding FD coefficients $A_{ij}$, $B_{ij}$, $C_{ij}$, and $D_{ij}$, respectively. The asterisk symbol ``$\ast$'' denotes TD convolution.
Causality is inherently enforced in these transformations, implying that the metasurface response at any given instant depends solely on present and past values of the incident fields. This physical constraint justifies setting the lower limit of the convolution integrals to zero.

\subsection{Time-Domain MAS Implementation of the GSTC}

To solve the time-dependent system of convolutional integral equations (\ref{eq:integral_equations}), the Method of Auxiliary Sources (MAS) is employed, as illustrated in Fig.~\ref{fig:geometry_MAS}(b), following the implementation approach described in~\cite{ludwig2011source}. From this point onward, the analysis is restricted to a $\mathrm{TM}_z$ polarization incident field originating in region $R_1$ and impinging upon the metasurface. 

MAS introduces two equivalent source configurations: one representing the scattered field in $R_1$ due to sources in $R_2$, and the other representing the transmitted field in $R_2$ due to sources in $R_1$. To this end, two auxiliary surfaces are introduced, denoted as $C^{(1)}_{\mathrm{aux}}$ and $C^{(2)}_{\mathrm{aux}}$, on which flow $z$-directed electric (for $\mathrm{TM}_z$ polarization) and magnetic (for $\mathrm{TE}_z$ polarization) current densities. The inclusion of both current types is necessary because the metasurface generally exhibits bianisotropic behavior; thus, the generated fields in each region contain both $\mathrm{TM}_z$ and $\mathrm{TE}_z$ components regardless of the incident polarization. The scattered field in $R_1$ is generated by the current densities on $C^{(1)}_{\mathrm{aux}}$, assuming the entire space has the material properties of $R_1$, while the transmitted field in $R_2$ arises from the densities on $C^{(2)}_{\mathrm{aux}}$ when the space has the properties of $R_2$. The fields radiated by the equivalent current densities on the auxiliary surfaces are expressed by \cite{ludwig2011source} (in case of $\mathrm{TM}_z$ polarization with \( J_z^{(q)} \) the $z$-directed \emph{electric} current density on \( C^{(q)}_{\mathrm{aux}} \))
\begin{subequations}
\label{eq:integrals_electric}
\begin{multline}
E^{(q)}_z(\boldsymbol{\rho}_q, t) = \\
-\frac{\eta_q}{2\pi}
\int_{C^{(q)}_{\mathrm{aux}}}
\int_{-\infty}^{t - P_q / c_q}
\frac{1}{\sqrt{c_q^2 (t - t')^2 - P_q^2}}
\frac{\partial J^{(q)}_z(\boldsymbol{\rho}'_q, t')}{\partial t'} 
\, \mathrm{d}t' \, \mathrm{d}l'_q,
\end{multline}
\begin{multline}
\boldsymbol{H}^{(q)}(\boldsymbol{\rho}_q, t) =
\frac{c_q}{2\pi}
\int_{C^{(q)}_{\mathrm{aux}}}
\int_{-\infty}^{t - P_q / c_q}
\frac{1}{\sqrt{c_q^2 (t - t')^2 - P_q^2}} \\
\times
\left\{ 
\hat{\mathbf{z}} \times 
\frac{\hat{\mathbf{P}}_q}{P_q} (t - t')
\frac{\partial J^{(q)}_z(\boldsymbol{\rho}'_q, t')}{\partial t'}
\right\}
\, \mathrm{d}t' \, \mathrm{d}l'_q.
\end{multline}
\end{subequations}
\noindent In case of $\mathrm{TE}_z$ polarization, the radiated fields are 
described via the \emph{magnetic} current densities \( M_z^{(q)} \) by
\begin{subequations}
\label{eq:integrals_magnetic}
\begin{multline}
H^{(q)}_z(\boldsymbol{\rho}_q, t) = 
-\frac{1}{2\pi \eta_q}\times\\
\int_{C^{(q)}_{\mathrm{aux}}}
\int_{-\infty}^{t - P_q / c_q}
\frac{1}{\sqrt{c_q^2 (t - t')^2 - P_q^2}}
\frac{\partial M^{(q)}_z(\boldsymbol{\rho}'_q, t')}{\partial t'} 
\, \mathrm{d}t' \, \mathrm{d}l'_q,
\end{multline}
\begin{multline}
\boldsymbol{E}^{(q)}(\boldsymbol{\rho}_q, t) =
-\frac{c_q}{2\pi}
\int_{C^{(q)}_{\mathrm{aux}}}
\int_{-\infty}^{t - P_q / c_q}
\frac{1}{\sqrt{c_q^2 (t - t')^2 - P_q^2}} \\
\times
\left\{ 
\hat{\mathbf{z}} \times 
\frac{\hat{\mathbf{P}}_q}{P_q} (t - t')
\frac{\partial M^{(q)}_z(\boldsymbol{\rho}'_q, t')}{\partial t'}
\right\}
\, \mathrm{d}t' \, \mathrm{d}l'_q,
\end{multline}
\end{subequations}
where $q = 1$ $(q=2)$ refers to $R_1$ and $C^{(1)}_{\mathrm{aux}}$ ($R_2$ and $C^{(2)}_{\mathrm{aux}}$), while \( c_q \) and \( \eta_q \) denote the wave velocity and intrinsic impedance of \( R_q \). Vectors \( \boldsymbol{\rho}_q \) and \( \boldsymbol{\rho}'_q \) specify observation and source points, with \( P_q = |\boldsymbol{\rho}_q - \boldsymbol{\rho}'_q| \)  and \( \hat{\mathbf{P}}_q = (\boldsymbol{\rho}_q - \boldsymbol{\rho}'_q)/P_q \).

As evident from~(\ref{eq:integrals_electric})–(\ref{eq:integrals_magnetic}), the evaluation of the fields in each region requires determining the unknown auxiliary current density derivatives. Hence, the next step in the MAS procedure involves discretizing the fictitious current density derivatives together with establishing a set of boundary (matching) points where the TD-GSTC will be enforced. The surface current densities are represented as superpositions of \( N_q \) filamentary sources, each positioned at a discrete location \( \boldsymbol{\rho}'_{qn} \) on \( C^{(q)}_{\mathrm{aux}} \). Furthermore, a set of \( M \) pairs of matching points is introduced on the metasurface boundary $C$ where the GSTCs are applied.

For \( \mathrm{TM}_z \) polarized reflected and transmitted fields, the auxiliary sources correspond to infinitely long electric current filaments oriented parallel to the $z$-axis, whereas for \( \mathrm{TE}_z \) polarized fields, they are replaced by equivalent magnetic current filaments aligned along the same axis. Applying a straightforward temporal discretization, we get 
\begin{equation}
\label{eq:dens_deriv_TM}
\frac{\partial J^{(q)}_z(\boldsymbol{\rho}'_q, t')}{\partial t'} =
\sum_{n} \sum_{k'}
I_{qn}^{(k')} \,
\delta_s(\boldsymbol{\rho}'_q - \boldsymbol{\rho}'_{qn}) \,
T_{qn}^{(k')}(t'),
\end{equation}
%
%
\begin{equation}
\label{eq:dens_deriv_TE}
\frac{\partial M^{(q)}_z(\boldsymbol{\rho}'_q, t')}{\partial t'} =
\sum_{n} \sum_{k'}
V_{qn}^{(k')} \,
\delta_s(\boldsymbol{\rho}'_q - \boldsymbol{\rho}'_{qn}) \,
T_{qn}^{(k')}(t'),
\end{equation}
\noindent
where \( I_{qn}^{(k')} \) and \( V_{qn}^{(k')} \) are the unknown amplitudes of the time derivatives of the electric and magnetic current densities, respectively, evaluated at the \( k' \)-th retarded time step on the \( n \)-th filamentary source located at position \( \boldsymbol{\rho}'_{qn} \). 

Function \( \delta_s(\boldsymbol{\rho}'_q - \boldsymbol{\rho}'_{qn}) \) is the surface delta function ensuring spatial localization of the source at the discrete filament position. The temporal basis function \( T_{qn}^{(k')}(t') \) defines the TD weighting of each source contribution and is chosen here as the simplest (piecewise constant) function
\begin{equation}
T_{qn}^{(k')}(t') =
\begin{cases}
1, & k'\Delta t < t' + \Delta T_{qn} < (k'+1)\Delta t, \\[4pt]
0, & \text{otherwise,}
\end{cases}
\label{eq:temporal_basis}
\end{equation}
\noindent
where \( \Delta t \) is the uniform time-step size. 
This representation enables the field integrals to be expressed as discrete time convolutions, which are later assembled into the matrix system that updates the fields at each time step in MAS–GSTC. Here,
\begin{equation}
\Delta T_{qn}= P^{(q)}_{\tilde m_n,n}/c_q
\label{eq:time_shift}
\end{equation}
is a source-dependent temporal shift, accounting for the propagation delay between each auxiliary source element and its nearest observation point, with
\begin{equation}
P^{(q)}_{mn}= |\boldsymbol \rho_{m}-\boldsymbol\rho'_{qn}|
\label{eq:distance}
\end{equation}
the Euclidean distance between the \( n \)-th source point on the \( q \)-th auxiliary curve and the \( m \)-th matching (observation) point on the physical curve \( C \). 
The index \( \tilde{m}_n \) identifies the observation (testing) point that is geometrically closest to the \( n \)-th source on the \( q \)-th auxiliary curve, ensuring that 
\begin{equation}
P^{(q)}_{\tilde{m}_n,n} = |\boldsymbol \rho_{\tilde{m}_n} - \boldsymbol \rho'_{qn}|
\end{equation}
corresponds to the minimum distance between the two. 
This choice guarantees that the temporal offset \( \Delta T_{qn} \) represents the smallest possible propagation delay between the \( n \)-th auxiliary source and its corresponding testing point, thereby improving the stability and accuracy of the TD marching-on-in-time process. 
Since the auxiliary sources are positioned on surfaces displaced from the physical boundaries, the retarded temporal discretization begins at 
\( t_q' = -P^{(q)}_{\tilde{m}_n,n}/c_q \). 
This negative retarded time ensures that the fields radiated by the auxiliary sources reach the physical surface exactly when the incident wavefront arrives, i.e., at \( t = 0 \). 

Substituting (\ref{eq:dens_deriv_TM})-(\ref{eq:dens_deriv_TE}) into (\ref{eq:integrals_electric})-(\ref{eq:integrals_magnetic}) and sampling the resulting equations at discrete time instants \( t_i = (i+1)\Delta t \) over the set of \( M \) matching points \( \boldsymbol{\rho}_{qm} \), we obtain, for each time step, the following discrete expressions for the electric field components
\begin{multline}
{E}_z^{(q)}(\boldsymbol{\rho}_{m}, (i+1)\Delta t)
= -\frac{\eta_q}{2\pi c_q}\\ 
\sum_{n}^{}
\Biggl(
\sum_{k'=0}^{i-\lfloor \kappa^{(q)}_{m,n} \rfloor - 1}
I_{qn}^{(k')}
\big[
W^{E^{(q)}}_{m,n}(i - k') -
W^{E^{(q)}}_{m,n}(i - k' - 1)
\big] \\
+~ I_{qn}^{(i-\lfloor \kappa^{(q)}_{m,n}\rfloor)} 
W^{E^{(q)}}_{m,n}(\lfloor \kappa^{(q)}_{m,n} \rfloor)
\Biggr),
\quad m = 1,2,\ldots,M
\label{eq:discrete_electric}
\end{multline}
where \( \kappa^{(q)}_{m,n} = (P^{(q)}_{m,n} - P^{(q)}_{\tilde{m}_n,n}) / (c_q \Delta t) \)
and \( \lfloor \kappa^{(q)}_{m,n} \rfloor \) denotes the greatest integer less than or equal to \( \kappa^{(q)}_{m,n} \), while
\begin{equation}
W^{E^{(q)}}_{m,n}(x)
= \operatorname{acosh}\!\left(
\frac{c_q \Delta t}{P^{(q)}_{m,n}}
\left[
x + \frac{P^{(q)}_{\tilde{m}_n,n}}{c_q \Delta t}
\right] + 1
\right).
\end{equation}

\noindent
Similarly, the magnetic field in each region is computed as
\begin{multline}
\boldsymbol{H}^{(q)}(\boldsymbol{\rho}_{m}, (i+1)\Delta t)
= \frac{1}{2\pi c_q}\\
\sum_{n}^{}
\Biggl(
\sum_{k'=0}^{i-\lfloor \kappa^{(q)}_{m,n} \rfloor - 1}
I_{qn}^{(k')}
\big[
\mathbf{W}^{H^{(q)}}_{m,n}(i - k') -
\mathbf{W}^{H^{(q)}}_{m,n}(i - k' - 1)
\big] \\
+~ I_{qn}^{(i-\lfloor \kappa^{(q)}_{m,n}\rfloor)}
\mathbf{W}^{H^{(q)}}_{m,n}(\lfloor \kappa^{(q)}_{m,n} \rfloor)
\Biggr),
\quad m = 1,2,\ldots,M
\label{eq:discrete_magnetic}
\end{multline}
where
\begin{equation}
\mathbf W_{m,n}^{H^{(q)}}(x)=\mathbf{\hat z}\times \mathbf{\hat P}_{m,n}\sqrt{\left( \frac{c_q\Delta t}{P^{(q)}_{m,n}}\left[x+\frac{P^{(q)}_{\tilde m_n,n}}{c_q\Delta t} +1\right]\right)^2-1 } 
\end{equation}
\noindent These equations provide the time-stepped evaluation of the scattered or transmitted fields in regions \( R_1 \) and \( R_2 \), respectively, as functions of the discretized auxiliary current amplitude derivatives. 
The same expressions are directly applicable to the \( \mathrm{TE}_z \) case through EM duality, by interchanging the roles of the electric and magnetic quantities.

To incorporate the TD GSTC into MAS, the convolution integrals appearing in (\ref{eq:integral_equations}) are discretized in time using a uniform time step $\Delta t$. 
As an example, the discretization of the first convolution term in the first equation of (\ref{eq:integral_equations}) is expressed as
\begin{multline}
a_{11}(t)\ast E_{1y}(t)
=\\ \sum_{k'=0}^{i} a_{11}\!\Bigl((i-k')\Delta t\Bigr)
E_{1y}\Bigl(\boldsymbol{\rho}_m,(k'+1)\Delta t\Bigr)\,\Delta t,
\label{eq:discrete_convolution}
\end{multline}
\noindent
with $a_{11}\Bigl((i-k')\Delta t\Bigr)$ the sampled impulse response associated with the corresponding surface susceptibility element, and $E_{1y}\Bigl(\boldsymbol{\rho}_m,(k'+1)\Delta t\Bigr)$ the tangential electric field evaluated at the $m$-th matching point and at time $(k'+1)\Delta t$. This discrete convolution accounts for the accumulated effect of all past field values on the present response, thus ensuring causal behavior of the metasurface system. A similar temporal discretization procedure is applied to all other convolution terms involving $a_{ij}(t)$, $b_{ij}(t)$, $c_{ij}(t)$, and $d_{ij}(t)$ in (\ref{eq:integral_equations}), yielding a fully discrete TD system suitable for numerical implementation within MAS. The final discretized GSTC is given in (\ref{eq:matrix}) for which,
\begin{multline}
\left[ \mathbf Z_{\scriptsize\left\{ \begin{smallmatrix} E^{q} \\ H^{q} \end{smallmatrix} \right\}}^{(j)} \right]_{m,n}
= \frac{1}{2\pi c_q}
\begin{Bmatrix}
-\eta_q \\
\Delta t
\end{Bmatrix}\\
\begin{cases}
W^{\scriptsize\left\{ \begin{smallmatrix} E^{q} \\ H^{q} \end{smallmatrix} \right\}}_{m,n}(j) - W^{\scriptsize\left\{ \begin{smallmatrix} E^{q} \\ H^{q} \end{smallmatrix} \right\}}_{m,n}(j-1), & j > \left\lfloor \kappa^{(q)}_{m,n} \right\rfloor \\
\quad W^{\scriptsize\left\{ \begin{smallmatrix} E^{q} \\ H^{q} \end{smallmatrix} \right\}}_{m,n}(j), & j = \left\lfloor \kappa^{(q)}_{m,n} \right\rfloor \\
\quad 0, & \text{otherwise}
\end{cases}
\end{multline}
\begin{align}
\boldsymbol{I}_q^{(j)} &=
\begin{bmatrix}
I_{qj}^{(0)} \\
I_{qj}^{(1)} \\
\vdots \\
I_{qj}^{(N)}
\end{bmatrix}, \quad
\boldsymbol{K}_q^{(j)} =
\begin{bmatrix}
K_{qj}^{(0)} \\
K_{qj}^{(1)} \\
\vdots \\
K_{qj}^{(N)}
\end{bmatrix}, \nonumber \\[4pt]
\mathbf{V}_E^{(j)} &= \Delta t
\begin{bmatrix}
E_z^{\mathrm{inc}}(\boldsymbol{\rho}_{1}, (j+1)\Delta t) \\
E_z^{\mathrm{inc}}(\boldsymbol{\rho}_{2}, (j+1)\Delta t) \\
\vdots \\
E_z^{\mathrm{inc}}(\boldsymbol{\rho}_{M}, (j+1)\Delta t)
\end{bmatrix}, \nonumber \\[4pt]
\mathbf{V}_H^{(j)} &= \Delta t
\begin{bmatrix}
H_y^{\mathrm{inc}}(\boldsymbol{\rho}_{1}, (j+1)\Delta t) \\
H_y^{\mathrm{inc}}(\boldsymbol{\rho}_{2}, (j+1)\Delta t) \\
\vdots \\
H_y^{\mathrm{inc}}(\boldsymbol{\rho}_{M}, (j+1)\Delta t)
\end{bmatrix}, \quad j = 0, 1, \ldots
\label{eq:vector_definitions}
\end{align}

\noindent The solution of (\ref{eq:matrix}) leads to the computation of the amplitudes $I_{qn}^{(k')}$, and subsequently to the calculation of the fields in each region using  (\ref{eq:discrete_electric}) and (\ref{eq:discrete_magnetic}) for observation points in $R_1$ and $R_2$ (instead of boundary points as the equations imply).

Before proceeding to the numerical results, it is useful to comment on the computational characteristics of the proposed TD--MAS formulation. The computational cost is mainly associated with the evaluation of the convolution terms appearing on the right-hand side of Eq.~\mbox{(\ref{eq:matrix})}. The system matrix corresponds to the zero-time-step GSTC matrix and remains unchanged throughout the marching-on-in-time procedure; therefore, its pseudo-inverse is computed only once at the beginning of the simulation. At each subsequent time step, only the right-hand side of the system is updated. In particular, the convolution terms include contributions from all previously computed current vectors, and, thus, additional history terms are incorporated as the simulation progresses. The increase in computational effort with simulation time, therefore, originates from evaluation of these accumulated convolution contributions rather than from repeated matrix inversion.

Regarding memory requirements, the impedance matrix depends only on the spatial discretization (number of auxiliary sources and matching points) and is independent of the total simulation duration. However, past current vectors must be stored to evaluate the convolution sums, and, thus, memory usage increases with the number of simulated time steps.

\section{Numerical Results and Discussion}
\label{sec:Numerical}
This section presents a comprehensive set of numerical experiments that demonstrate the accuracy, stability, and versatility of the TD--MAS formulation combined with GSTC. We begin with a detailed workflow example involving an anisotropic (magnetostatically biased) graphene layer, which also serves to validate the method through error analysis and stability testing. This case highlights the full simulation procedure—from parameter selection to field evaluation in each region—and establishes practical implementation guidelines. Subsequent sections examine a black phosphorus metasurface, which also exhibits anisotropic EM behavior, and an artificial metasurface characterized by Lorentzian surface susceptibilities. In all simulations, the parameters $N_1$, $N_2$, $M$, and $\Delta t$ collectively define the \emph{spatiotemporal resolution} of the TD--MAS--GSTC method, as they determine the spatial sampling of the auxiliary sources and matching points, as well as the temporal marching accuracy.

\subsection{Magnetostatically Biased Anisotropic Graphene Sheet Excited by a Current Filament}
\label{sec:graphene_case}

Although the anisotropic graphene layer is not a metasurface in the conventional sense, it is nevertheless well suited as a test scenario for the proposed MAS--GSTC formulation. The considered graphene layer 
is fully characterized by an anisotropic surface conductivity tensor of the form
\begin{equation}
\overline{\overline{\sigma}} =
\begin{bmatrix}
\sigma_{yy} & \sigma_{yz} \\
\sigma_{zy} & \sigma_{zz}
\end{bmatrix}
\label{eq:graphene_sigma_tensor}
\end{equation}
where the off-diagonal terms $\sigma_{yz}$ and $\sigma_{zy}$ arise due to the Hall effect when a static magnetic bias $B_0$ is applied normal to the graphene sheet. The components of the tensor in the FD are
\begin{align}
\sigma_{yy}(\omega, B_0) &= \sigma_{zz}(\omega, B_0) =
\frac{\sigma_0 (1 + \mathrm{i} \, \omega \tau)}
{(\omega_c \tau)^2 + (1 + \mathrm{i} \, \omega \tau)^2}, \label{eq:sigma_diag}\\
\sigma_{zy}(\omega, B_0) &= - \sigma_{yz}(\omega, B_0) =
\frac{\sigma_0 \, \omega_c \tau}
{(\omega_c \tau)^2 + (1 + \mathrm{i} \, \omega \tau)^2}, \label{eq:sigma_offdiag}
\end{align}
with $\sigma_0$ and $\omega_c$ the DC conductivity and cyclotron frequency
\begin{equation}
\sigma_0 = \frac{2 e^2 \tau}{\pi \hbar^2} k_B T 
\ln \!\left( 2 \cosh \frac{E_F}{2 k_B T} \right), 
\quad
\omega_c = \frac{e B_0 v_F^2}{E_F}.
\label{eq:sigma_params}
\end{equation}

The surface electric susceptibility tensor entering the GSTC is related to the conductivity tensor through
\begin{equation}
\overline{\overline{\chi}}_{ee}(\omega, B_0)
= \frac{\mathrm{i}}{\omega \varepsilon_0}
\, \overline{\overline{\sigma}}(\omega, B_0),
\label{eq:chi_sigma_relation}
\end{equation}
leading to
\[
\overline{\overline{\chi}}_{ee} =
\begin{bmatrix}
 \chi^{yy}_{ee} & \chi^{yz}_{ee} \\
\chi^{zy}_{ee} & \chi^{zz}_{ee}
\end{bmatrix}.
\]
Since graphene does not exhibit magnetic behavior, the corresponding magnetic susceptibility components are absent. This means that only the FD elements $A_{ij}$ and their TD correspondents $a_{ij}(t)$ are not zero, even though the general MAS--GSTC formulation presented in Section~\ref{sec:theory} accounts for both electric and magnetic surface responses. For the simulations we have: $e = 1.62 \times 10^{-19}~\mathrm{C}$ the elementary charge, $\tau = 0.6~\mathrm{ps}$~\cite{Zouaghi2015} the electron scattering time, $\hbar = 1.054 \times 10^{-34}~\mathrm{J \cdot s}$ the reduced Planck constant, $k_B = 1.380 \times 10^{-23}~\mathrm{J/K}$ the Boltzmann constant, $T = 300~\mathrm{K}$ the temperature, $E_F = 0.5~\mathrm{eV}$ the Fermi energy, $v_F = 10^6~\mathrm{m/s}$ the Fermi velocity, and $ B_0=1~\mathrm{T}$ the $x$-directed external magnetic bias applied to the graphene sheet.

The convolution kernels $a_{ij}(t)$ required by the TD--MAS formulation are obtained by applying IFT to (\mbox{\ref{eq:sigma_diag}}) and (\mbox{\ref{eq:sigma_offdiag}}). The resulting TD conductivities are given by
\begin{subequations}
\begin{equation}
\sigma_{yy}(t)=\sigma_{zz}(t)=\frac{\sigma_0}{\tau} e^{-t/\tau}\cos(\omega_c t),
\end{equation}
\begin{equation}
\sigma_{yz}(t)=-\sigma_{zy}(t)=-\frac{\sigma_0}{\tau} e^{-t/\tau}\sin(\omega_c t),
\end{equation}
\end{subequations}
\noindent These kernels are causal and exponentially decaying, and therefore no singular or impulsive contributions arise. In the numerical implementation, they are sampled directly at the marching time instants $t=k\Delta t$ and used in the discrete convolution sums without truncation or windowing. Due to their exponential decay, contributions from sufficiently old time samples naturally become negligible in the simulation durations considered. 

The incident field is a transient $\rm TM_z$ cylindrical wave generated by a current filament. Due to the anisotropic nature of magnetically biased graphene, reflected and transmitted fields generally include also $\rm TE_z$ components, resulting in polarization conversion in addition to scattering. The source current is a differentiated Gaussian pulse with amplitude in Amperes,
\begin{equation}
I_0(t) = - \left( \frac{t}{\tau_p} \right) \exp \left[ 0.5 - 0.5 \left( \frac{t}{\tau_p} \right)^2 \right],
\label{eq:Gaussian-pulse}
\end{equation}
where $\tau_p=3\times 10^{-14}~s$ denotes the pulse width parameter, introduced here to distinguish it from the electron scattering time $\tau$ of graphene. 
The position of the source is set to $(x_s, y_s)=(-c_0\tau_p,0)=(-9~\rm \mu m,~0)$ with $c_0$ the speed of light in vacuum. The Nyquist sampling interval for this differentiated Gaussian pulse is $\Delta t_0 = 1/(2 f_0)$, with $f_0$ the effective upper frequency bound of the pulse spectrum. This frequency is chosen such that the spectral amplitude satisfies
\begin{equation}
I_0(f = f_0) = 10^{-4} \, I_0(f = 0).
\end{equation}
Based on this criterion, the cutoff frequency and sampling interval are approximated as $f_0 \approx 1.932 / \tau_p$ and $\Delta t_0 \approx \tau_p / 3.864$, respectively. For the present setup, this gives $\tau_p = 3 \times 10^{-14}~\mathrm{s}$, $f_0 = 64.355~\mathrm{THz}$, and $\Delta t_0 \approx 7.769 \times 10^{-15}~\mathrm{s}$. This frequency content ensures that only the intraband term of graphene's susceptibility is relevant, consistent with the expressions given in (\ref{eq:sigma_diag})-(\ref{eq:sigma_offdiag}), which consider solely the intraband contribution.

\begin{figure}[htb!]
\centering
\subfigure[]{\includegraphics[width=0.24\textwidth]{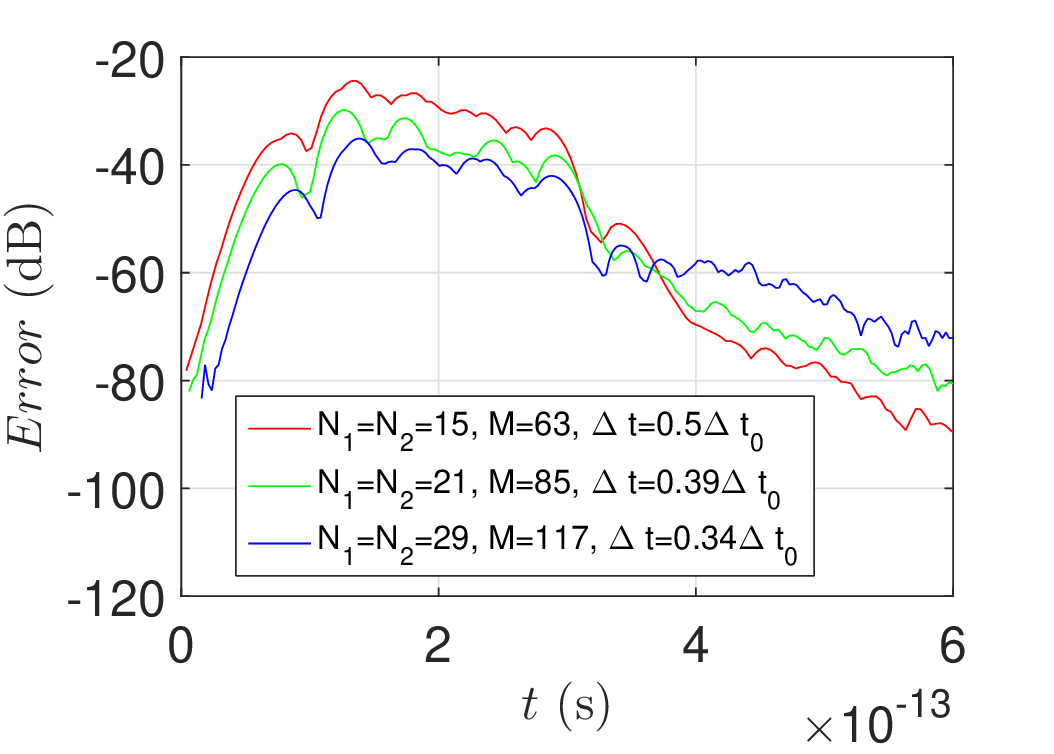}} 
\hfill
    \subfigure[]{\includegraphics[width=0.24\textwidth]{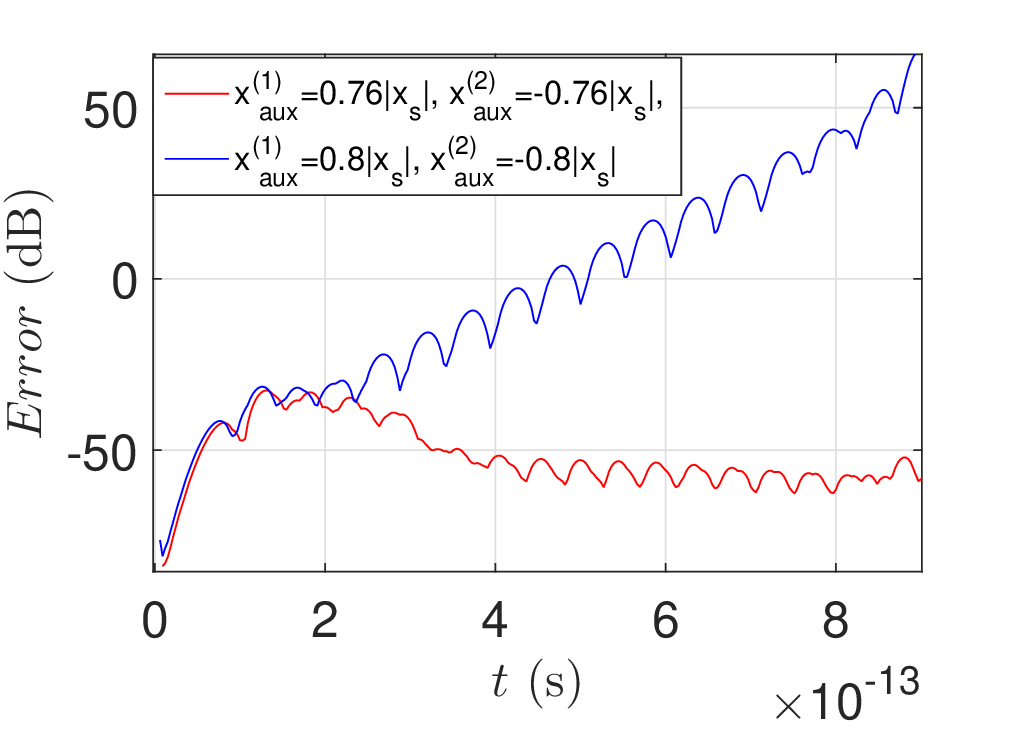}}
    \hfill  
    \subfigure[]{\includegraphics[width=0.24\textwidth]{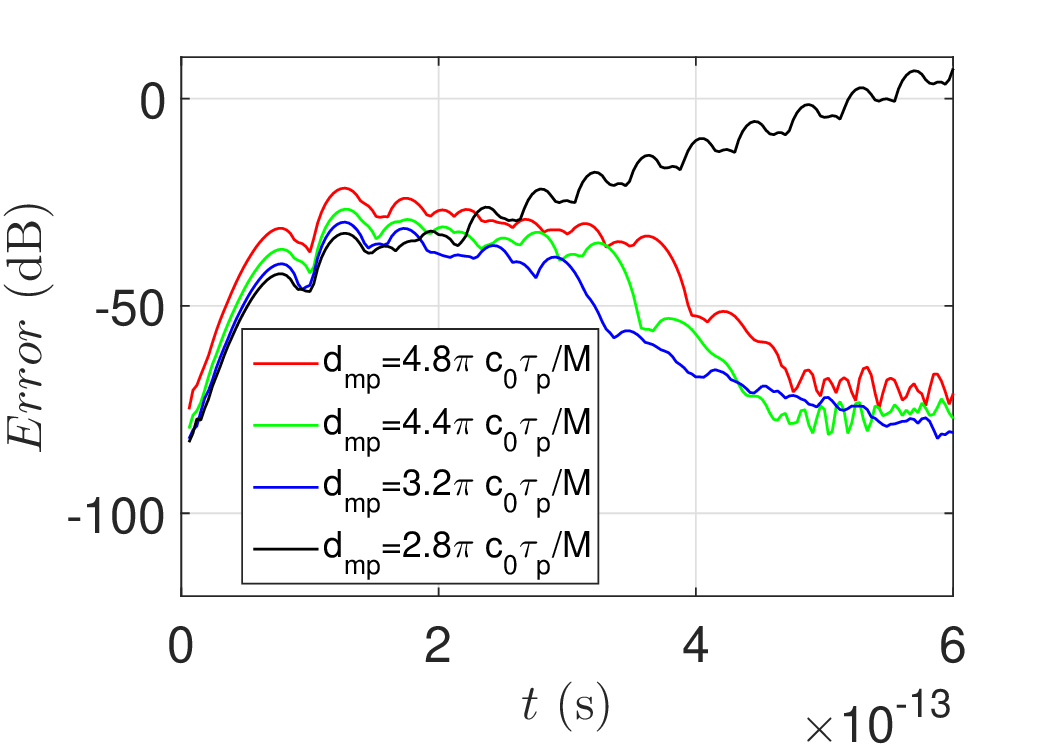}}
    \hfill  
\caption{(a) Temporal evolution of the boundary error for three different spatio-temporal resolutions $(N_1, N_2, M, \Delta t)$. 
    (b) Effect of auxiliary surface placement on numerical stability for $N_1 = N_2 = 21$, $M = 85$, and $\Delta t = 0.39\,\Delta t_0$.(c) Effect of $d_{mp}$ on the boundary error for $N_1 = N_2 = 21$, $M = 85$.}
\label{fig:conv_stab}\unskip
\end{figure}

\paragraph{Boundary Error and Stability}
To verify the accuracy and temporal robustness of the proposed formulation, we first examine the boundary condition error defined in (\ref{eq:boundary_error}) in which all field quantities are evaluated at $(y,t)$, and $\ast$ denotes convolution
with respect to $t$ with the modified kernels
\begin{align*}
\tilde b_{12}(t) &= b_{12}(t)-\delta(t), &
\tilde b'_{12}(t) &= b_{12}(t)+\delta(t),\\
\tilde b_{21}(t) &= b_{21}(t)+\delta(t), &
\tilde b'_{21}(t) &= b_{21}(t)-\delta(t),\\
\tilde d_{12}(t) &= d_{12}(t)+\delta(t), &
\tilde d'_{12}(t) &= d_{11}(t)-\delta(t),\\
\tilde d_{21}(t) &= d_{21}(t)+\delta(t), &
\tilde d'_{21}(t) &= d_{21}(t)-\delta(t),
\end{align*} 

\noindent where $\delta(t)$ is the delta function. Figure~\ref{fig:conv_stab}(a) illustrates the temporal evolution of this error for two combinations of \( (N_1, N_2, M, \Delta t) \). 
Finer spatial and temporal resolutions yield noticeably improved convergence behavior, confirming the method’s numerical consistency.


Although the TD MAS--GSTC scheme exhibits adequate accuracy, it is not strictly unconditionally stable. At long simulation times, numerical instabilities may emerge once the accumulated boundary error exceeds a tolerable level, defining a maximum stable duration \( T_{\mathrm{max}} \). In the present work, $T_{\mathrm{max}}$ is defined as the largest time instant for which the boundary error (\mbox{\ref{eq:boundary_error}}) remains below $-25$~dB, i.e.,
\begin{equation*}
T_{\rm max} = \max \left\{ t \;:\; \mathrm{Error}(t) \leq -25~\mathrm{dB} \right\}.
\end{equation*}
Beyond this time instant, the accumulated marching-on-in-time and convolutional contributions lead to rapid growth of the boundary error, indicating numerical instability.
As shown in Fig.~\ref{fig:conv_stab}(b), this stability limit can be substantially extended by positioning the auxiliary surfaces closer to the physical interface. 
This adjustment improves numerical conditioning and delays the onset of divergence without any increase in computational cost.

To further quantify the dependence of the stability limit on auxiliary-surface placement and source position, we performed a systematic parametric study under the fixed spatiotemporal resolution $N_1 = N_2 = 21$, $M = 85$, and $\Delta t = 0.39\,\Delta t_0$. The resulting values of $T_{\rm max}$ are summarized in Table~\mbox{\ref{tab:stability_results}}. It is observed that for moderate source distances 
($x_s=-c_0\tau_p$ and $x_s=-2c_0\tau_p$), positioning the auxiliary surfaces within the range 
$0.7c_0\tau_p \le |x_{\rm aux}| < 0.8c_0\tau_p$ significantly extends the stability duration. 
Larger auxiliary distances (e.g., $|x_{\rm aux}|=0.84c_0\tau_p$) lead to a reduction in $T_{\rm max}$. 
For very small source distances ($x_s=-c_0\tau_p/5$), improved stability is achieved for auxiliary surfaces placed approximately at the same distance as the source (i.e., $|x_{\rm aux}| \approx |x_s|$), together with a slight reduction in $d_{mp}$. 
These observations provide practical guidance for auxiliary-surface positions maximizing stability.

\begin{table}[htbp]
\centering
\caption{$T_{\rm max}$ values with respect to the position of the auxiliary surfaces for three different positions of the excitation source}
\label{tab:stability_results}
\begin{tabular}{c c c c}
\hline
$x_s$ & $(x^{(1)}_{\rm aux},x^{(2)}_{\rm aux})$ & $d_{mp}$ & $T_{\max}$ \\
\hline
\multirow{3}{*}{$-c_0\tau_p$}
& $(-0.84c_0\tau_p,\; +0.84c_0\tau_p)$ & $3.2\pi c_0\tau_p/M$ & $92\Delta t$ \\
& $(-0.80c_0\tau_p,\; +0.80c_0\tau_p)$ & $3.2\pi c_0\tau_p/M$ & $120\Delta t$ \\
& $(-0.70c_0\tau_p,\; +0.70c_0\tau_p)$ & $3.2\pi c_0\tau_p/M$ & $>300\Delta t$ \\
\hline
\multirow{3}{*}{$-2c_0\tau_p$}
& $(-0.84c_0\tau_p,\; +0.84c_0\tau_p)$ & $3.2\pi c_0\tau_p/M$ & $94\Delta t$ \\
& $(-0.80c_0\tau_p,\; +0.80c_0\tau_p)$ & $3.2\pi c_0\tau_p/M$ & $126\Delta t$ \\
& $(-0.70c_0\tau_p,\; +0.70c_0\tau_p)$ & $3.2\pi c_0\tau_p/M$ & $>300\Delta t$ \\
\hline
$-\dfrac{c_0\tau_p}{5}$
& $\left(-\dfrac{c_0\tau_p}{5},\; +\dfrac{c_0\tau_p}{5}\right)$
& $2.6\pi c_0\tau_p/M$ & $>300\Delta t$ \\
\hline
\end{tabular}
\end{table}

\begin{figure*}[b]
\centering
\resizebox{0.9\textwidth}{!}{%
\begin{minipage}{\textwidth}
\begin{multline}
\mathrm{Error~(dB)} =\\
20\log_{10}\left[
\frac{
\underset{\boldsymbol{y} \in C}{\max}\left|
\begin{bmatrix}
a_{11}\!\ast E_{1y}+\tilde b'_{12}\!\ast H_{1z} &
a_{11}\!\ast E_{2y}+ \tilde b_{12}\!\ast H_{2z} &
a_{12}\!\ast E^{sc}_{1z}+b_{11}\!\ast H^{sc}_{1y} &
a_{12}\!\ast E_{2z}+b_{11}\!\ast H_{2y}
\\[4pt]
a_{21}\!\ast E_{1y}+b_{22}\!\ast H_{1z} &
a_{21}\!\ast E_{2y}+b_{22}\!\ast H_{2z} &
a_{22}\!\ast E^{sc}_{1z}+\tilde b_{21}\!\ast H^{sc}_{1y} &
a_{22}\!\ast E_{2z}+\tilde b'_{21}\!\ast H_{2y}
\\[4pt]
d_{11}\!\ast E_{1y}+c_{12}\!\ast H_{1z} &
d_{11}\!\ast E_{2y}+c_{12}\!\ast H_{2z} &
\tilde d_{12}\!\ast E^{sc}_{1z}+c_{11}\!\ast H_{1y} &
\tilde d'_{12}\!\ast E_{2z}+c_{11}\!\ast H_{2y}
\\[4pt]
\tilde d'_{21}\!\ast E_{1y}+c_{22}\!\ast H_{1z} &
\tilde d_{21}\!\ast E_{2y}+c_{22}\!\ast H_{2z} &
d_{22}\!\ast E^{sc}_{1z}+c_{21}\!\ast H^{sc}_{1y} &
d_{22}\!\ast E_{2z}+c_{21}\!\ast H_{2y}
\end{bmatrix}
\right|
}{
\underset{t}{\max}\left(
\underset{\boldsymbol{y}\in C}{\max}\left|
\begin{bmatrix}
 -a_{12}\!\ast E^{inc}_{z} - b_{11}\!\ast H^{inc}_{y}\\[4pt]
 -a_{22}\!\ast E^{inc}_{z} - \tilde b_{21}\!\ast H^{inc}_{y}\\[4pt]
 -\tilde d_{12}\!\ast E^{inc}_{z} - c_{11}\!\ast H^{inc}_{y}\\[4pt]
 -d_{22}\!\ast E^{inc}_{z} - c_{21}\!\ast H^{inc}_{y}
\end{bmatrix}
\right|
\right)
}
\right]
\label{eq:boundary_error}
\end{multline}
\end{minipage}
} 
\end{figure*}

It should be noted that unlike grid-based TD methods, the proposed TD--MAS formulation does not follow a Courant--Friedrichs--Lewy (CFL) stability condition that limits the time step according to the spatial discretization. Instead, instability in TD--MAS may appear after long simulation times due to the gradual accumulation of numerical errors during the marching-on-in-time procedure and the contribution of past field values through the convolution terms. As a result, although TD--MAS--GSTC is not unconditionally stable in the strict CFL sense, the stable simulation time can be extended by properly positioning the auxiliary surfaces and selecting suitable discretization parameters, as demonstrated by Table~\mbox{\ref{tab:stability_results}}.

\paragraph{Implementation Guidelines}
Based on the preceding analysis, the following workflow is proposed for setting up simulations (it applies also to the black phosphorus case)
\begin{enumerate}
    \item Specify $\tau_p$ such that the frequency content of the incident pulse (in the THz region) is appropriate to justify the use of only the intraband term of graphene's conductivity.
    \item Place the excitation source at any point along the $x$-axis.
    \item Start with \( N_1 = N_2 = 15 \) and \( M =4 N_1+3 \).
    \item The auxiliary surfaces are initially placed at $x^{(1)}_{\rm aux} = 0.8c_0\tau_p$, $x^{(2)}_{\rm aux} = -0.8c_0\tau_p$, and must always satisfy $ |x_{\rm aux}| < |x_s|$ to ensure causal consistency (otherwise the incident field reaches the matching points before the scattered contribution). If the source is placed such that $|x_s|\le |x^{(1)}_{\rm aux}|$, then both auxiliary-surface positions are reduced so that $|x^{(1)}_{\rm aux}|,|x^{(2)}_{\rm aux}|<|x_s|$.

\item Set the spacing between successive matching points to $d_{mp}=3.2 \pi c_0\tau_p / M$, and that between successive source points to $d_{\rm aux}=(M/N) d_{mp}$. When $|x_s|\le c_0\tau_p/4$, a slight reduction of $d_{mp}$ improves the boundary error.
    \item If the boundary error exceeds the desired tolerance (e.g., above $-25~\rm dB$), increase the number of sources and matching points following the empirical rule $N_1=N_2,~M=4N_1+3$, and reduce the time step \( \Delta t \).   
    \item When time reaches \( T_{\rm max} \), move the auxiliary surfaces closer to the physical boundary to enhance stability.
\end{enumerate}

At this point, it is useful to clarify the choice of the matching-point spacing $d_{mp}$, as its specific value may appear arbitrary at first glance. The selection $d_{mp}=3.2\pi c_0\tau_p/M$ is not ad hoc. An initial value $d_{mp}=2\pi c_0\tau_p/M$ was considered based on analogous TD--MAS formulations for cylindrical scatterers \mbox{\cite{TDSIBCarxiv, TD_shielding}}, where $2\pi c_0\tau_p$ corresponds to the circumference of a circle of radius $c_0\tau_p$ and provides stable sampling when matching points are uniformly distributed along that contour. However, for the planar metasurface configuration examined here, this spacing was found to reduce stability due to the different geometric distribution of retarded-time contributions. Systematic numerical experiments showed that increasing the spacing to $d_{mp}=3.2\pi c_0\tau_p/M$ (or slightly reducing it when the source is placed very close to the metasurface) improves the boundary condition error and significantly extends $T_{\rm max}$. This value therefore represents a practical compromise between stability and accuracy for the planar TD--MAS--GSTC formulation.

To further assess the robustness of the recommended $d_{mp}$ (and essentially $d_{\rm aux}$ due to their relation  $d_{\rm aux}=(M/N)d_{mp}$), we performed a concise sensitivity study by varying $d_{mp}$ around the proposed value, while keeping all other discretization parameters fixed. In addition to the baseline choice $
d_{mp}=3.2\pi c_0\tau_p/M$ the values $d_{mp}=4.8\pi c_0\tau_p/M,\; 4.4\pi c_0\tau_p/M,\; 2.8\pi c_0\tau_p/M$ were examined. The resulting boundary-error evolution is shown in Fig.~\mbox{\ref{fig:conv_stab}(c).} It is observed that increasing $d_{mp}$ above the recommended value preserves stability over the considered time window but leads to noticeably increased boundary residual. Conversely, reducing $d_{mp}$ below the recommended value significantly decreases $T_{\rm max}$, indicating reduced stability margin. These results confirm that the selected value $d_{mp}=3.2\pi c_0\tau_p/M$ lies within a stable operating window that balances boundary-condition error and stability.

Although the numerical experiments for extracting the above stability and discretization rules were performed for the anisotropic graphene case, the same geometric considerations apply to the black phosphorus configuration and, more generally, to any metasurface excited by a localized point source. In all such cases, the auxiliary-surface placement must satisfy the same ordering condition relative to the source position, and the matching-point spacing rule remains unchanged. For the Lorentzian metasurface, where the excitation is a Gaussian-pulsed plane beam rather than a localized filament, the same principles for auxiliary-surface positioning and matching-point spacing apply. Since the excitation is not associated to a single localized source position, the auxiliary placement is selected relative to the characteristic spatial scale $c_0\tau_p$ of the pulse, and no additional source-distance constraint is required.

\paragraph{Final Parameters and Field Results}

Based on the above process, the final discretization parameters are
\begin{enumerate}
    \item Position of the source: $x_s=-c_0\tau_p=-9~\rm \mu m$
    \item Auxiliary surfaces: $x^{(1)}_{\rm aux} = 0.76 |x_s|$, $x^{(2)}_{\rm aux} = -0.76 |x_s|.$
    \item Number of sources: \( N_1 = 21 \), \( N_2 = 21 \)
    \item Matching points: \( M = 85 \)
    \item Distance between successive source and matching point: $d_{\rm aux}=(M/N)d_{mp}$, with $d_{mp}\approx 1.064~\rm{\mu m}$.  
    \item Time step: \( \Delta t = 0.4 \Delta t_0 \)
\end{enumerate}

\begin{figure}[htb!]
\centering
\subfigure[]{\includegraphics[width=0.24\textwidth]{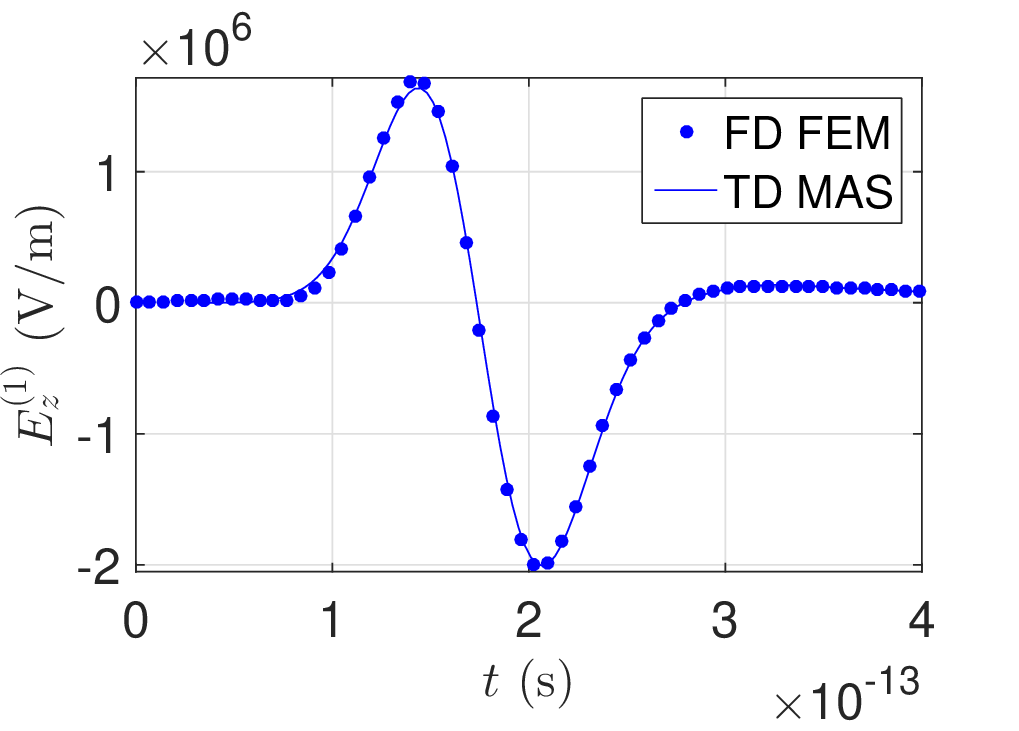}} 
\hfill
    \subfigure[]{\includegraphics[width=0.24\textwidth]{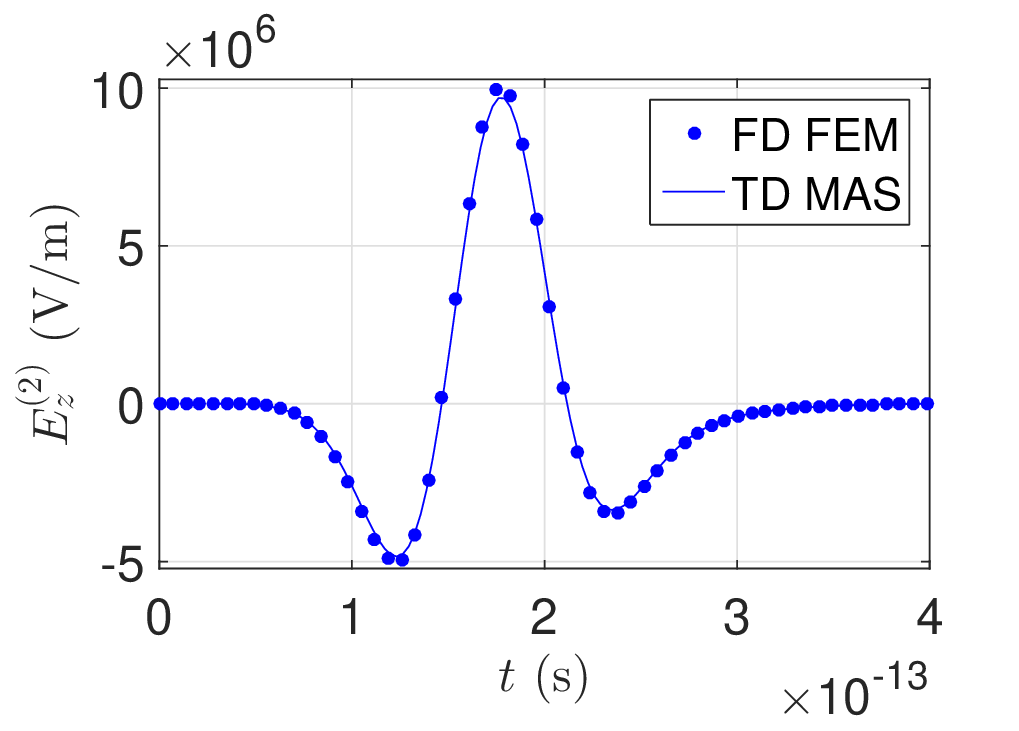}}
    \hfill
    \subfigure[]{\includegraphics[width=0.243\textwidth]{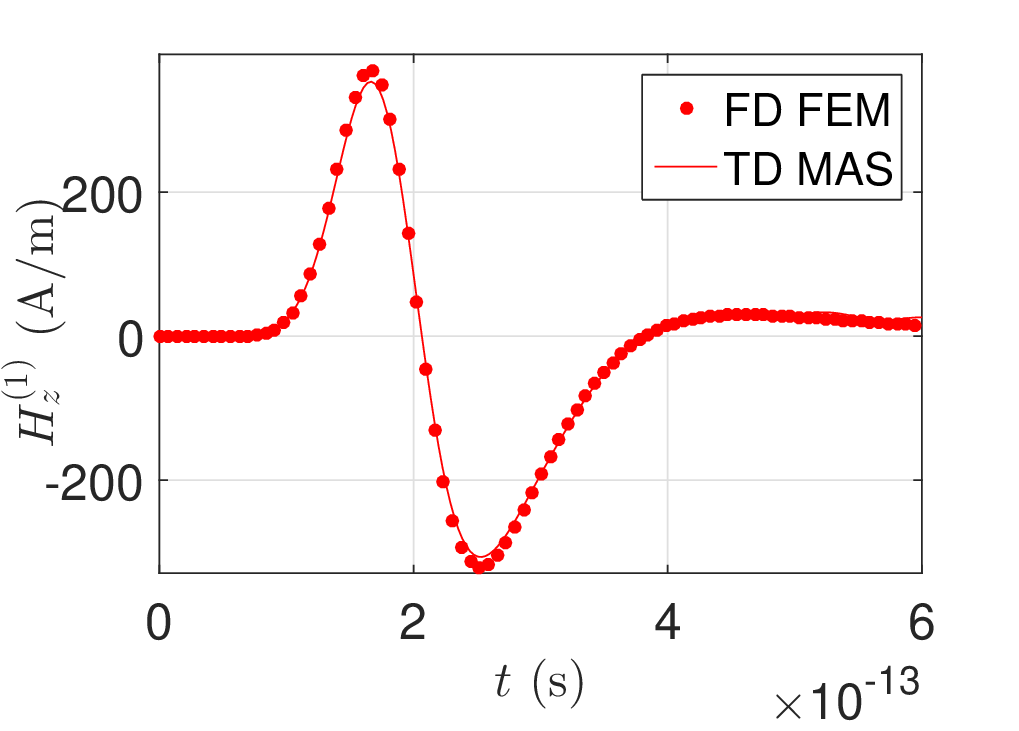}} 
\hfill
    \subfigure[]{\includegraphics[width=0.24\textwidth]{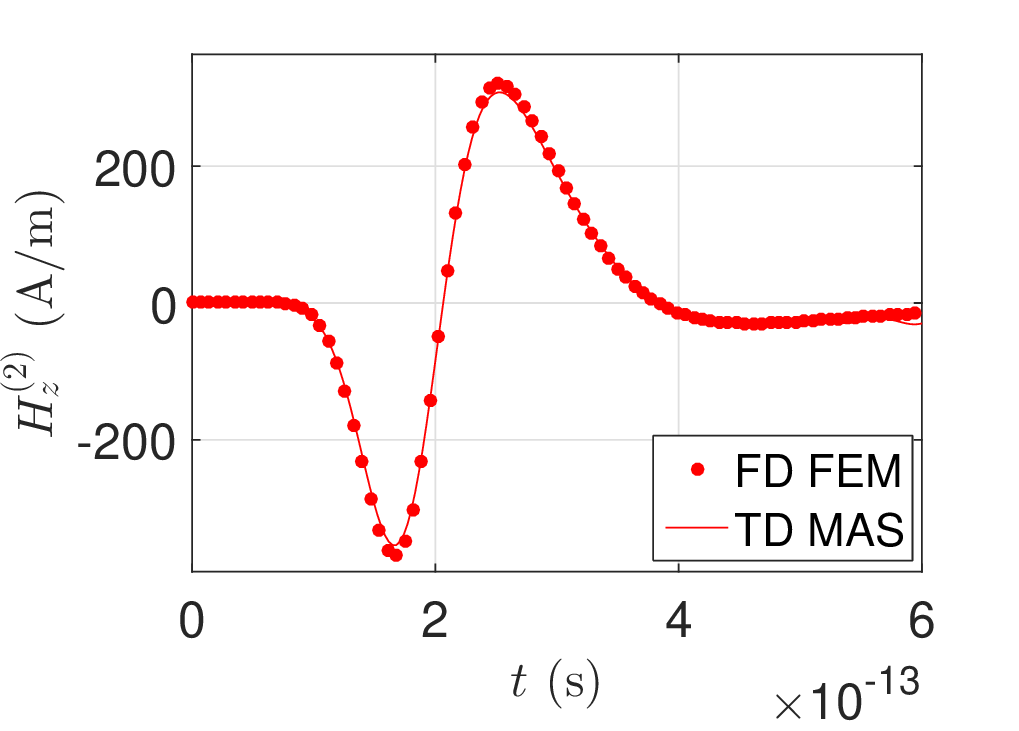}}
    \hfill
    \subfigure[]{\includegraphics[width=0.243\textwidth]{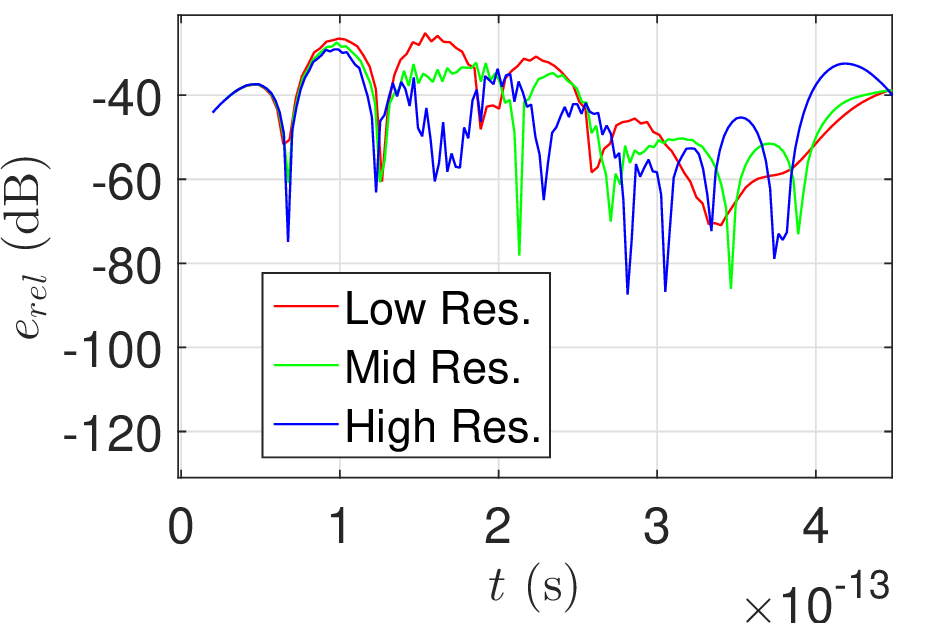}} 
\hfill
    \subfigure[]{\includegraphics[width=0.24\textwidth]{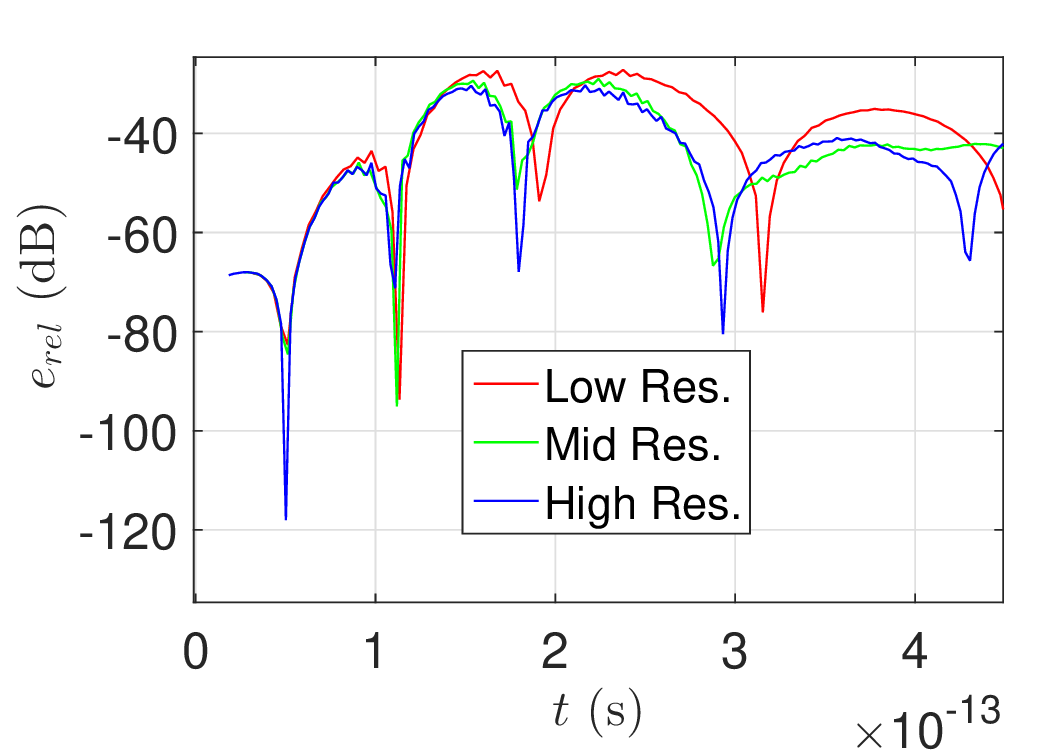}}

\caption{Transient $z$-directed field components for the anisotropic graphene metasurface: 
    (a) reflected $E_z^{(1)}$, (b) transmitted $E_z^{(2)}$, 
    (c) reflected $H_z^{(1)}$, and (d) transmitted $H_z^{(2)}$. 
    Fields are sampled at $(x_1, y_1)=(-0.1|x_s|,0)$ in region~$R_1$ and 
    $(x_2, y_2)=(0.1|x_s|,0)$ in region~$R_2$. Instantaneous relative field error $e_{rel}$ for the reflected (e) electric $E^{(1)}_z$ and (f) magnetic $H_z^{(1)}$ fields.} 
\label{fig:graphene}\unskip
\end{figure}

With these choices, the method provides accurate and stable results. 
Validation of the MAS results was carried out using 2145 FD samples computed with COMSOL Multiphysics over the range $33.333~\mathrm{GHz} \le f \le 71.506~\mathrm{THz}$, followed by application of the IFFT on the samples. It is noted that since the COMSOL version employed did not implement GSTCs, the graphene layer was modeled as a surface current boundary condition, taking into account its anisotropic surface conductivity tensor. For the intraband graphene model considered here, the surface current satisfies $\mathbf{J}=\overline{\overline{\sigma}}\,\mathbf{E}_t$ (where $\mathbf{E}_t$ the total tangential electric field on the boundary), and since the conductivity tensor is related to the electric surface susceptibility through $\overline{\overline{\sigma}}=\mathrm{i}\omega\varepsilon_0\overline{\overline{\chi}}_{ee}$, this surface-current representation is equivalent to the electric-type GSTC formulation used in the TD--MAS model.
Figs.~\ref{fig:graphene}(a), (b) and Figs. ~\ref{fig:graphene}(c), (d) show the scattered $z$-directed electric and magnetic fields in the two regions, $E^{(1)}_z$ ($H^{(1)}_z$) and $E^{(2)}_z$ ($H^{(2)}_z$), demonstrating that the MAS–GSTC method achieves good agreement with the COMSOL results.

\paragraph{Resolution Sensitivity Study}

To examine the influence of spatio-temporal resolution on numerical accuracy, we performed a systematic refinement study for the graphene metasurface example. Three resolution levels were considered:
\begin{itemize}
\item Low resolution: $N_1=N_2=15$, $M=63$, $\Delta t=0.5\Delta t_0$,
\item Mid resolution: $N_1=N_2=21$, $M=85$, $\Delta t=0.39\Delta t_0$,
\item High resolution: $N_1=N_2=29$, $M=117$, $\Delta t=0.34\Delta t_0$,
\end{itemize}
\noindent while keeping the auxiliary-surface placement consistent with the implementation guidelines.

Fig.~\mbox{~\ref{fig:conv_stab}}(a) shows the boundary error evolution for the three resolutions. Reduction in boundary error is observed as the spatial and temporal discretization is refined. In particular, the high-resolution configuration exhibits both improved boundary-condition enforcement and delayed onset of instability. To further quantify accuracy, the instantaneous relative field error (in dB) between the TD--MAS solution and the reference solution is computed for both the reflected $E_z$ and $H_z$ components. The relative electric field error is defined as
\begin{equation}
e_{\mathrm{rel}}(t)~(\mathrm {dB})
=
20\log_{10}\frac{|E_{\mathrm{TD\text{-}MAS}}(t)-E_{\mathrm{ref}}(t)|}
{\max |E_{\mathrm{ref}}(t)|},
\label{eq:relative_error}
\end{equation}
\noindent where $E_{\mathrm{TD\text{-}MAS}}$ is the TD-MAS computed electric field and $E_{\mathrm{ref}}$ is the reference field which in the anisotropic graphene case is the field by FD-COMSOL combined with IFFT. An analogous relative error can be defined for the magnetic field. The plots in \mbox{Figs.~\ref{fig:graphene}}~(e) and (f) 
demonstrate a generally improved performance of the error with increasing resolution.

\subsection{Anisotropic Black Phosphorus Sheet Excited by a Current Filament}
Black phosphorus (BP) is another 2-D material that exhibits in-plane anisotropy. Its EM response is characterized by a diagonal electric surface conductivity tensor, where the two principal components correspond to the so-called ``armchair'' ($y$-direction) and ``zigzag'' ($z$-direction) crystal orientations. These components generally take different values, rendering BP intrinsically anisotropic. The anisotropic surface conductivity of BP is described by a semiclassical Drude model as
\begin{equation}
\sigma_{s,jj}(\omega) = \frac{D_j}{\pi \left( \mathrm{i} \omega + \frac{\eta_e}{\hbar} \right)}, \quad j \in \{y, z\},
\label{eq:Drude}
\end{equation}
where $\eta_e$ is the electron relaxation rate, and $D_j = \pi e^2 n_{sj} / m_j$ denotes the Drude weight. Here, $n_{sj}$ is the electron surface carrier concentration and $m_j$ represents the electron effective mass along the $j$-th in-plane direction. The effective masses along the $y$ (armchair) and $z$ (zigzag) directions are
\begin{equation}
m_y = \frac{\hbar^2}{\frac{2\gamma^2}{\Delta_{\mathrm{BP}}} + \eta_c}, 
\quad
m_z = \frac{\hbar^2}{2 v_c},
\end{equation}
where $\gamma = 4d / \pi$, $\eta_c = \hbar^2 / (0.4 m_0)$, and $v_c = \hbar^2 / (1.4 m_0)$. 
In these expressions, $\Delta_{\mathrm{BP}}$ is the thickness-dependent band-gap of black phosphorus, $d$ is the thickness of the layer, and $m_0=9.109\times 10^{-31}~\rm kg$ denotes the electron rest mass. Here, the parameters are set to \cite{jang2022efficient} $\Delta_{\mathrm{BP}} = 2~\mathrm{eV}$, $\eta_e = 10~\mathrm{meV}$, $n_{sj} = 10^{13}~\mathrm{cm^{-2}}$, and $d = 0.223~\mathrm{nm}$. The corresponding electric surface susceptibilities are then obtained via
\begin{equation}
\chi^{jj}_{ee}(\omega) = \frac{\sigma_{s,jj}(\omega)}{\mathrm{i} \omega \varepsilon_0}, \quad j \in \{y, z\}.
\end{equation}
\noindent This means that only the diagonal FD elements $A_{jj}$ and their TD correspondents $a_{jj}(t)$ are not zero.

The convolution kernels $a_{jj}(t)$ are obtained by applying IFT to the FD expressions of the surface conductivity tensor associated with the semiclassical Drude model of black phosphorus (\mbox{\ref{eq:Drude}}). This yields the exact TD conductivities
\begin{subequations}
\begin{equation}
\sigma_{yy}(t)=\frac{D_y}{\pi}\exp\!\left(-\frac{\eta_e}{\hbar}t\right),
\end{equation}
\begin{equation}
\sigma_{zz}(t)=\frac{D_z}{\pi}\exp\!\left(-\frac{\eta_e}{\hbar}t\right).
\end{equation}
\end{subequations}

\noindent As in the graphene case, the kernels are sampled at the marching time instants and employed directly in the discrete convolution sums without truncation or windowing.

The anisotropic BP layer is illuminated by a TD electric current filament at $(x_s, y_s) = (-57~\mathrm{\mu m}, 0)$. The incident field generated by this source corresponds to a transient $\rm TM_z$ cylindrical wave. The source current is expressed by the same differentiated Gaussian pulse as in the graphene case, given by (\ref{eq:Gaussian-pulse}), with
%
%
$\tau_p = x_s / c_0$. For the present configuration, this yields $\tau_p \approx 1.901 \times 10^{-13}~\mathrm{s}$, resulting in a sampling interval $\Delta t_0 = 4.920 \times 10^{-14}~\mathrm{s}$ and an effective cutoff frequency $f_0 \approx 10.161~\mathrm{THz}$. These parameters ensure that the excitation spectrum remains in the frequency range where the semiclassical Drude model accurately describes the EM behavior of BP.

According to the MAS--GSTC formulation described in Section~\ref{sec:theory}, two auxiliary surfaces, $C^{(1)}_{\rm aux}$ and $C^{(2)}_{\rm aux}$, are introduced on either side of the BP layer. Following the \hlg{same} workflow procedure as in the anisotropic graphene case,  the optimal positions of these surfaces were found to be
\begin{equation}
x^{(1)}_{\rm aux} = 0.76 |c_0\tau_p|, \quad x^{(2)}_{\rm aux} = -0.76 |c_0\tau_p|.
\end{equation}
Each auxiliary surface hosts $N_1 = N_2=21$ sources, while $M =85$ temporal matching points are employed with a time step $\Delta t=0.39 \Delta t_0$. The spacing between successive auxiliary sources on both surfaces is set to $d_{\rm aux} = (M/N)d_{mp}= 27~\mathrm{nm}$, and the separation between consecutive matching points is $d_{mp} = 6.67~\mathrm{nm}$. These parameters were determined to provide satisfactory numerical stability and an acceptably low boundary error over the simulation time window.

\begin{figure}[htb!]
\centering
\subfigure[]{\includegraphics[width=0.24\textwidth]{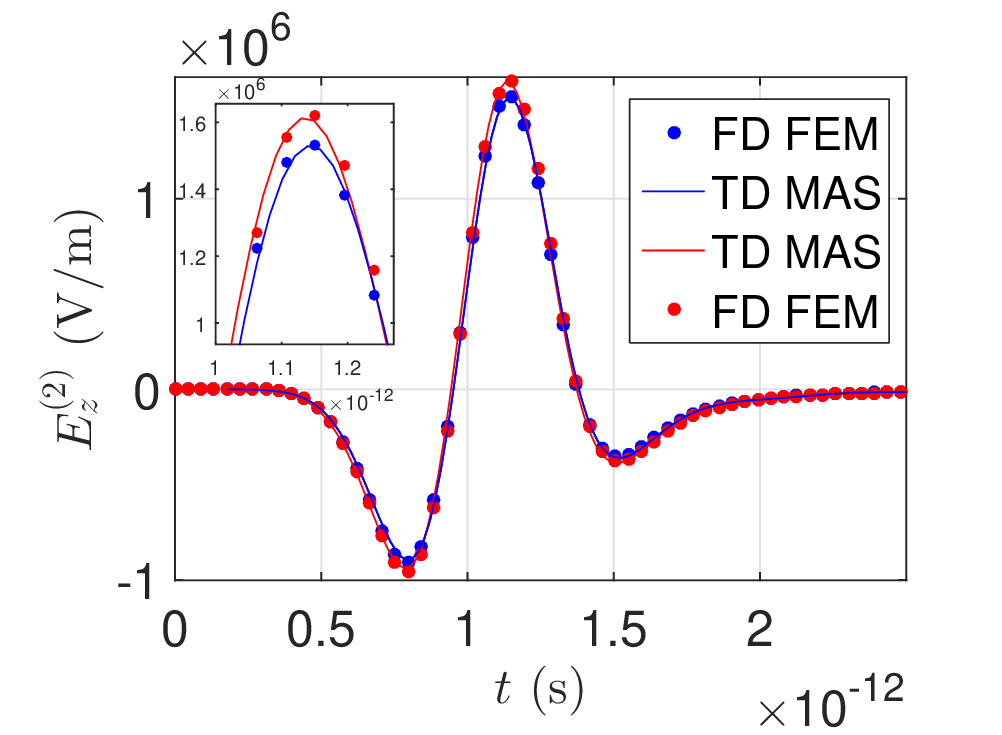}} 
\hfill
    \subfigure[]{\includegraphics[width=0.24\textwidth]{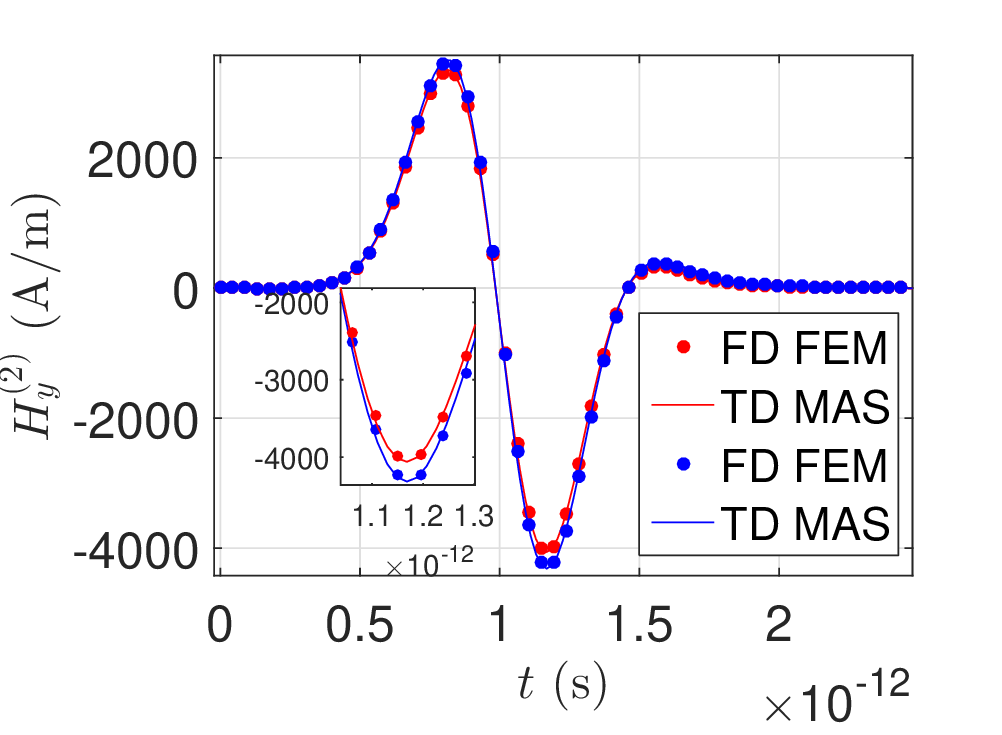}} 
    \hfill
    \subfigure[]{\includegraphics[width=0.24\textwidth]{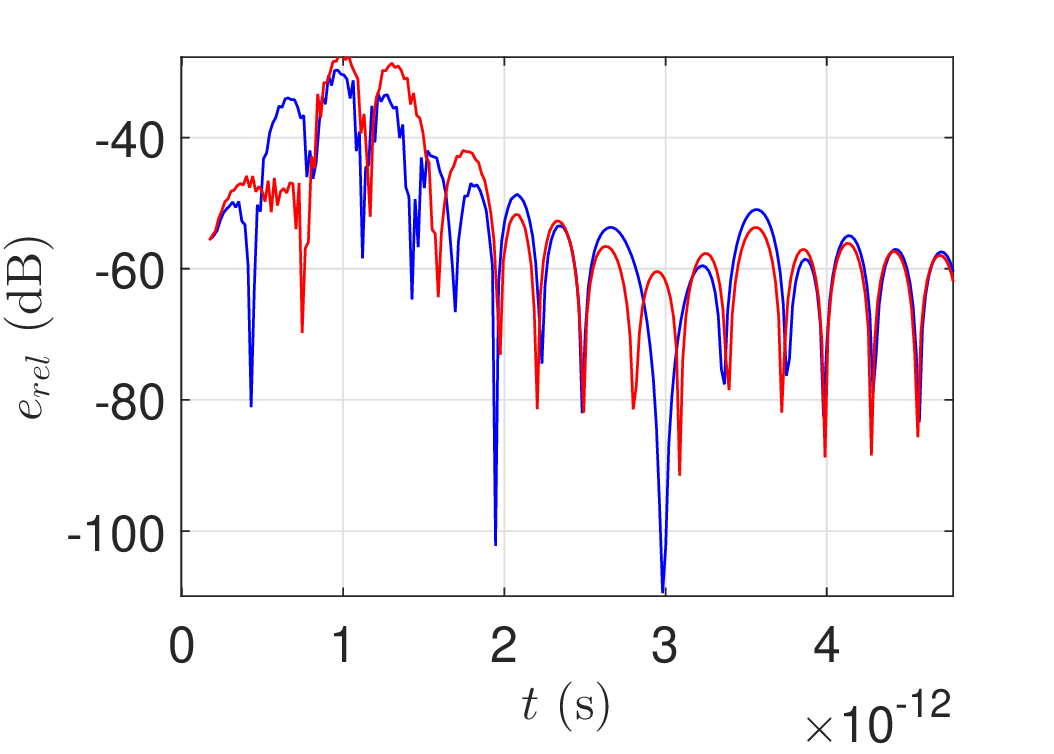}}
    \hfill
    \subfigure[]{\includegraphics[width=0.24\textwidth]{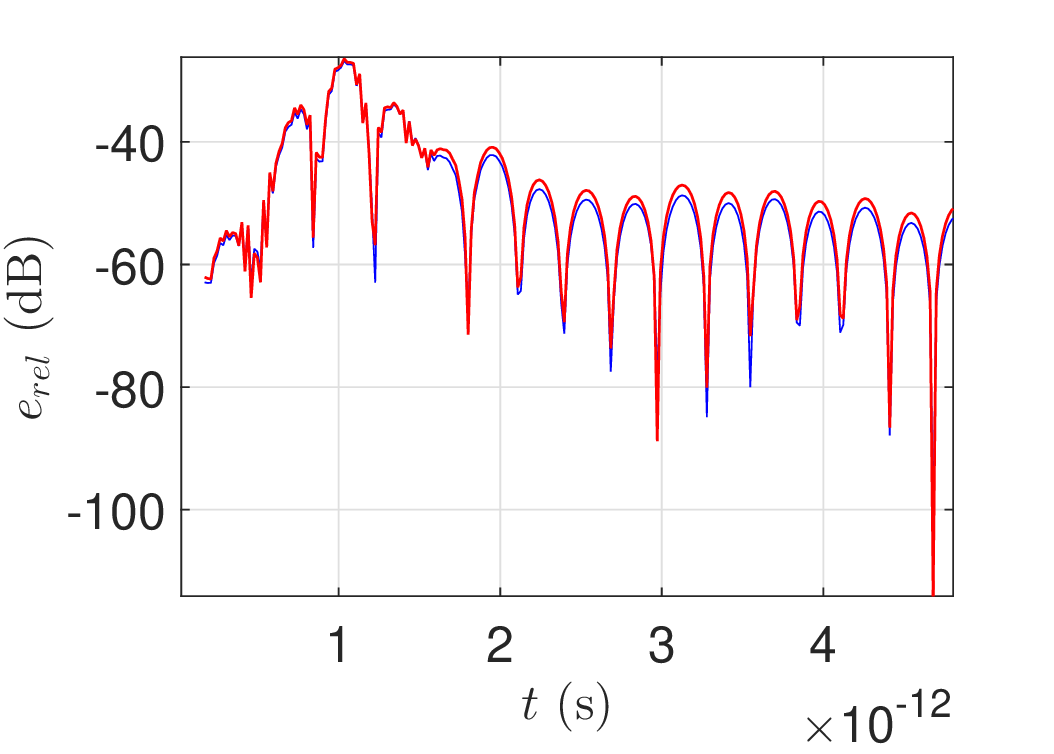}}
\caption{Transient field components for the anisotropic black phosphorus metasurface: 
    (a) $E_z$ and (b) $H_y$ in region $R_2$ (transmitted fields), 
    for both crystallographic orientations—armchair along $y$, zigzag along $z$ (blue dots and lines) and the reverse (red dots and lines). Inset plots show the small difference due to anisotropy. 
    Observation point: $(x_2, y_2)=(0.1|x_s|,0)$ in $R_2$. Instantaneous relative field error $e_{rel}$  for the: (c) transmitted electric field $E^{(2)}_z$ and (d) transmitted magnetic field $H_y^{(2)}$.} 
\label{fig:phosphorene}\unskip
\end{figure}

The results obtained from the MAS--GSTC implementation are presented in the Fig. \ref{fig:phosphorene}, where the transmitted $z$-directed electric field $E^{(2)}_z$ and the $y$-directed magnetic \hlg{field} $H^{(2)}_y$ are plotted for:  
(1) the ``armchair'' conductivity oriented along the $y$-direction and the ``zigzag'' conductivity along the $z$-direction, and  
(2) the ``armchair'' conductivity oriented along the $z$-direction and the ``zigzag'' conductivity along the $y$-direction.  
The observed differences between these two cases illustrate the in-plane anisotropy of BP. In addition, we plot in Figs. \mbox{\ref{fig:phosphorene}}(c) and (d) the relative error $e_{rel}$ (for both cases) defined in (\mbox{\ref{eq:relative_error}}) with reference solutions obtained by FD--COMSOL and IFFT. 

\subsection{Artificial Lorentzian Metasurface Excited by a Gaussian-Pulsed Plane Wave}
In this example, we consider an artificial metasurface characterized by Lorentzian-type electric and magnetic surface susceptibilities \cite{smy2020fdtd}. For simplicity, only two nonzero susceptibility components are assumed, namely $\chi^{(yy)}_{ee}$ and $\chi^{(zz)}_{mm}$, given by
\begin{align}
\label{eq:gupta_susc}
\chi^{(yy)}_{ee}(\omega) &= \frac{\omega_{ep}^2}{\omega_{e0}^2 - \omega^2 + \mathrm{i} a_e \omega}, \\
\chi^{(zz)}_{mm}(\omega) &= \frac{\omega_{mp}^2}{\omega_{m0}^2 - \omega^2 + \mathrm{i} a_m \omega}.
\end{align}
\noindent This means that only the FD elements $A_{11}$, $C_{22}$ of (\ref{eq:Rmatrix}), (\ref{eq:Lmatrix}) and their TD counterparts $a_{11}(t)$, $c_{22}(t)$ are not zero. 
In (\ref{eq:gupta_susc}) $\omega_p$, $\omega_0$, and $a$ are the plasma frequency, resonant frequency, and the loss factor respectively. The subscripts $e$ and $m$ denote electric and magnetic quantities. These expressions correspond to the standard Lorentzian dispersion model, which can, in general, include multiple resonant terms but is here reduced to a single resonance for each susceptibility since higher-order resonances are negligible within the considered frequency range.

The convolution kernels are extracted from the FD expressions of the surface susceptibilities via an IFT and given by
\begin{subequations}
\begin{multline}
\chi^{yy}_{ee}(t)=
\varepsilon_0\omega_{ep}^{2}
\exp\!\left(-\frac{\alpha_e}{2}t\right)
\\\frac{\left(\cos(\omega_{e0}t)
-\frac{\alpha_e}{2\omega_{e0}}\sin(\omega_{e0}t)\right)}{2},
\end{multline}
\begin{multline}
\chi^{zz}_{mm}(t)=
\mu_0\omega_{mp}^{2}
\exp\!\left(-\frac{\alpha_m}{2}t\right)
\\\frac{\left(\cos(\omega_{m0}t)
-\frac{\alpha_m}{2\omega_{m0}}\sin(\omega_{m0}t)\right)}{2}
\end{multline}
\end{subequations}

\noindent Also here, the kernels are sampled at the marching time instants and do not require truncation or windowing.

In the present numerical investigation, we assume that the plasma and loss parameters are identical for both electric and magnetic responses, i.e., $\omega_{mp} = \omega_{ep}$ and $a_m = a_e$. Two distinct scenarios are examined depending on the values of the resonant frequencies $\omega_{m0}$ and $\omega_{e0}$: (1) the \emph{matched case}, where $\omega_{m0} = \omega_{e0}$, leading to negligible reflection and nearly complete transmission through the metasurface, and (2) the \emph{mismatched case}, where $\omega_{m0} \neq \omega_{e0}$, resulting in both reflected and transmitted fields. 

Since such a Lorentzian metasurface realistically, exhibits a narrow-band response around its resonant frequency, it cannot be meaningfully excited by a broadband transient source, such as the differentiated Gaussian filament used above. Instead, a TD Gaussian-pulsed plane wave is employed to provide a controlled, band-limited excitation centered around a desired frequency. Specifically, we consider a normally incident $\rm TE_z$-polarized Gaussian-pulsed plane wave, centered at frequency $f_0$, with a temporal pulse width $\tau_p$ and a transverse beam width equal to the free-space wavelength $\lambda_0 = c_0 / f_0$. The incident transient magnetic field is defined as
\begin{equation}
H_z^{\mathrm{inc}}(x,y,t) = \exp \left( -\frac{y^2}{\lambda_0^2} \right)
\exp \left[ -\frac{(t - x / c_0)^2}{\tau_p^2} \right] \cos (\omega_0 t),
\end{equation}
where $\omega_0 = 2 \pi f_0$. This field represents a Gaussian-pulsed beam that propagates along the positive $x$-direction while attenuating transversely away from the $y$-axis.

The parameters used in this simulation are as follows: the central frequency is set to $f_0 = 230~\mathrm{THz}$, the pulse width to $\tau_p = 8 \times 10^{-15}~\mathrm{s}$, and the electric and magnetic loss factors are $a_e = a_m = 7.54 \times 10^{12}$. The plasma frequency is taken as $\omega_{ep} = \omega_{mp} = 3.01 \times 10^{11}~\mathrm{rad/s}$. For the \emph{matched case}, $\omega_{e0} = \omega_{m0} = 2\pi \times 230~\mathrm{THz}$, resulting in negligible reflection and nearly perfect transmission through the metasurface. In contrast, for the \emph{mismatched case}, $\omega_{e0} - \omega_{m0} = 2\pi \times 25~\mathrm{THz}$, which produces both reflected and transmitted fields.

In accordance with the MAS--GSTC framework, and by following a similar workflow with the previous cases, 
we have $x^{(1)}_{\mathrm{aux}} = 0.27 c_0\tau_p= 0.647 ~\mathrm{\mu m}$ and $x^{(2)}_{\mathrm{aux}} = 0.27 c_0\tau_p=-0.647~\mathrm{\mu m}$. The number of temporal matching points was set to $M = 81$, with a spacing of $d_{\rm mp}=1.2 \pi c_0\tau_p/M\approx 0.111 ~\mathrm{\mu m}$. Each auxiliary surface hosts $N = 21$ sources, spaced at intervals of $d_{\rm aux}=(M/N)d_{mp}$. Lastly the time step is set at $\Delta t\approx 0.26~\rm {fs}$. These parameters ensured satisfactory numerical stability and a sufficiently low boundary error, consistent with the procedure outlined in Section~\ref{sec:theory}.

\begin{figure}[htb!]
\centering
\subfigure[]{\includegraphics[width=0.24\textwidth]{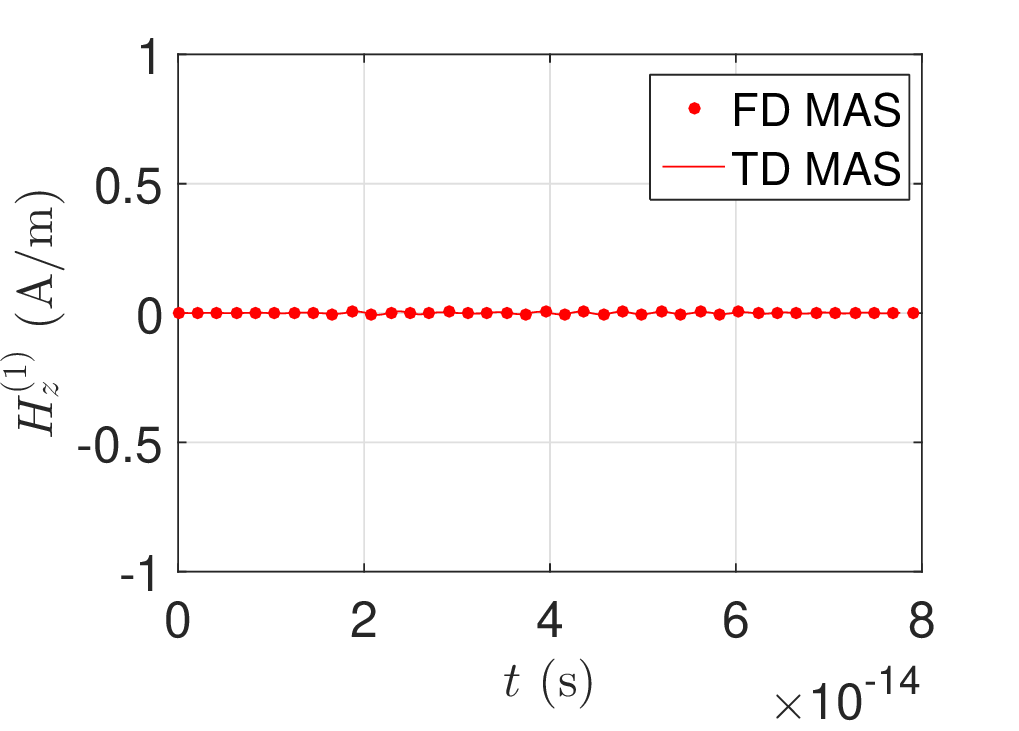}} 
\hfill
    \subfigure[]{\includegraphics[width=0.24\textwidth]{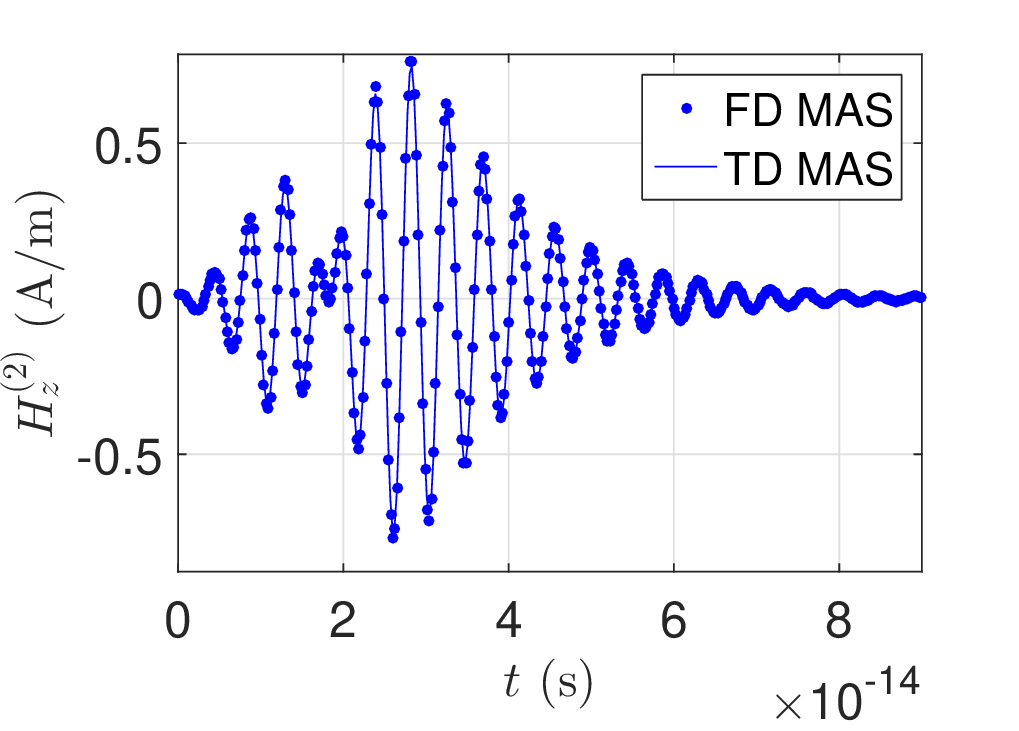}}
    \hfill
    \subfigure[]{\includegraphics[width=0.243\textwidth]{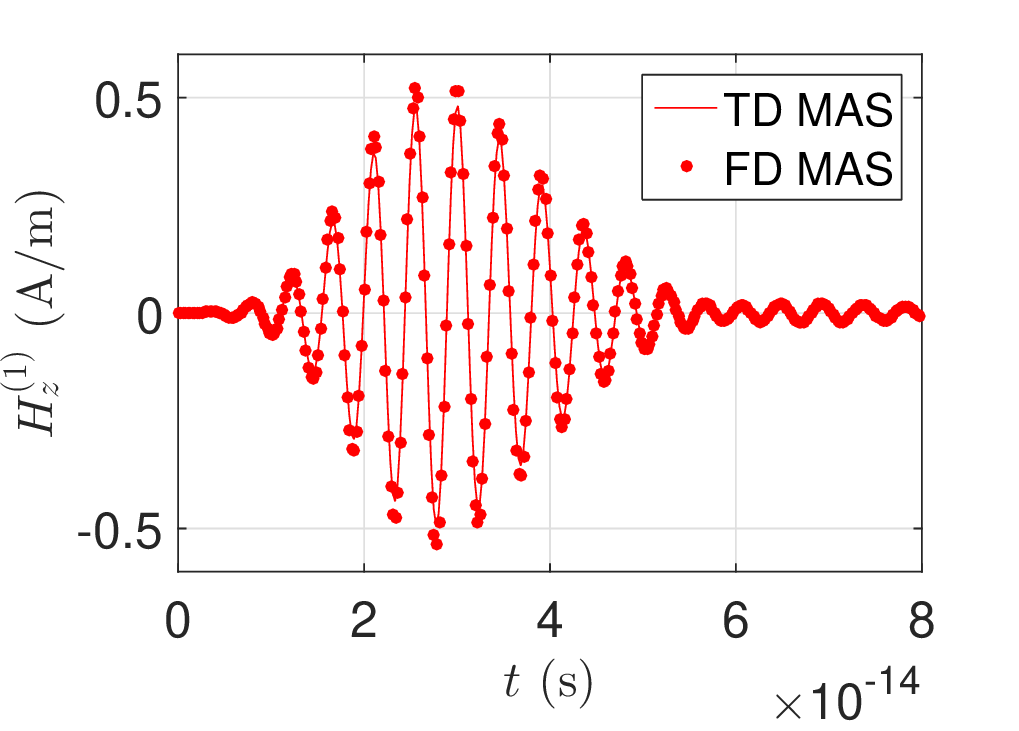}} 
\hfill
    \subfigure[]{\includegraphics[width=0.24\textwidth]{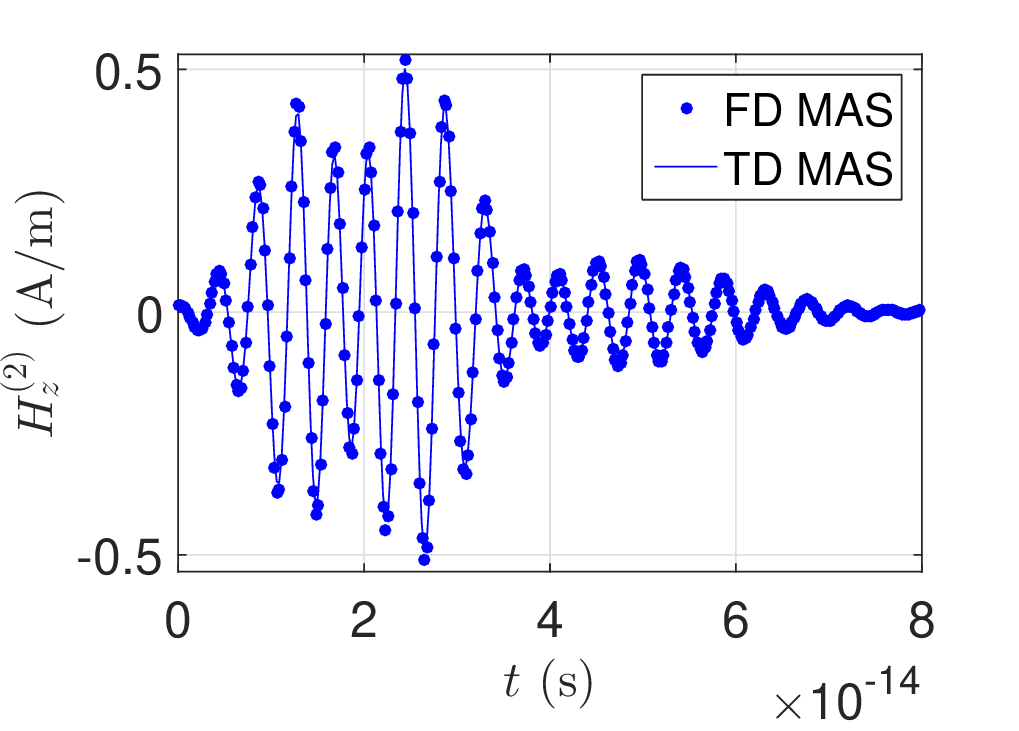}}
    \hfill
    \subfigure[]{\includegraphics[width=0.243\textwidth]{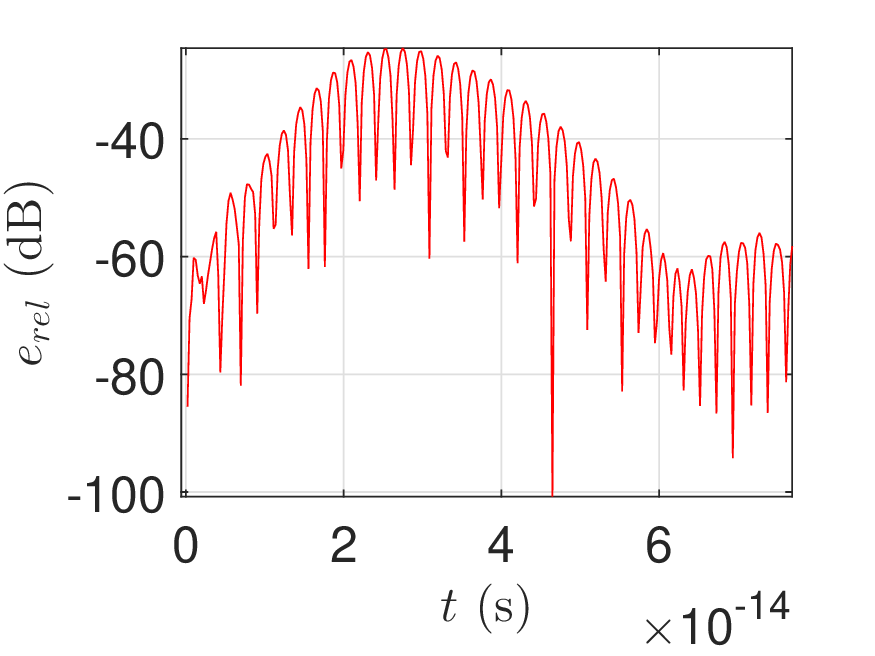}} 
\hfill
    \subfigure[]{\includegraphics[width=0.24\textwidth]{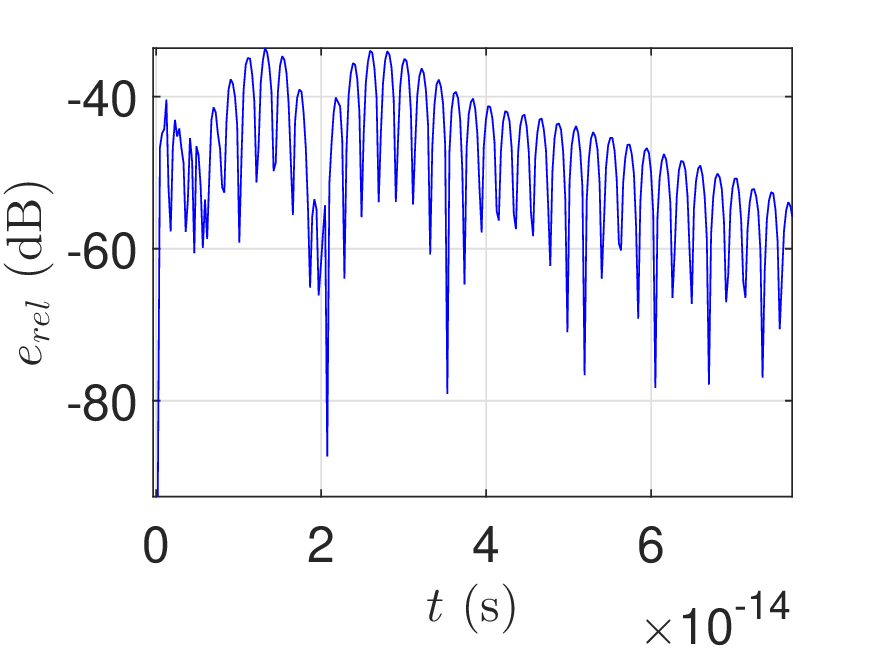}}
\caption{Time-domain $H_z$ field responses in regions $R_1$ and $R_2$ for the artificial Lorentzian metasurface. 
    (a)--(b) Matched configuration and (c)--(d) mismatched configuration. 
    Observation points: $(x_1, y_1)=(-0.1|c_0\tau_p|,0)$ in $R_1$ and $(x_2, y_2)=(0.1|c_0\tau_p|,0)$ in $R_2$. Instantaneous relative error $e_{rel}$ for the mismatched case for the: (e) reflected field $H^{(1)}_z$ and (f) transmitted field $H_z^{(2)}$.}
\label{fig:gupta}\unskip
\end{figure}

The TD results are presented in  Fig. \ref{fig:gupta}, illustrating the transmitted and reflected transient fields for the matched and mismatched cases. As expected, the matched configuration yields a negligible reflected field, while the mismatched configuration exhibits clear reflection alongside transmission. The MAS--GSTC results were validated against those of a FD MAS solver \cite{wang2020simulation}, followed by IFFT of 4089 frequency samples covering the spectral bandwidth $0.234~\rm THz-1923~THz$ which is much larger than that of the incident Gaussian pulse. Figs. \mbox{\ref{fig:gupta}}(e) and (f) show the relative error $e_{rel}$, defined in (\mbox{\ref{eq:relative_error}}), for the mismatched case with reference solutions by FD--MAS and IFFT.

\subsection{Computational Performance}

To evaluate the computational performance of the proposed TD--MAS method, we compare CPU time and peak memory usage with the reference solvers used for validation in this work. A direct comparison with grid-based TD methods is not considered, since stable FDTD or TD-FEM simulations of dispersive metasurfaces require specialized implementation and tuning, which was not available in the present study. Instead, performance is assessed here via comparisons with FD-COMSOL and FD-MAS simulations combined with IFFT of the results.

All simulations were performed on a workstation equipped with an Intel Xeon W-2133 processor (3.6\,GHz) and 48\,GB of RAM. For the graphene and black-phosphorus examples, the proposed TD--MAS implementation required approximately 5\,GB of RAM and about 50 minutes of CPU time to complete the full broadband transient simulation. For the same physical configurations, the corresponding FD--COMSOL simulations required approximately 2.4\,GB of RAM and about 110 minutes of CPU time. For the Lorentzian metasurface, comparison was carried out against the FD--MAS approach. The TD--MAS simulation required again 5\,GB of RAM and 50 minutes of CPU time, whereas the FD--MAS solution required about 0.5\,GB of RAM and approximately 105 minutes due to the need for multiple frequency-domain solutions followed by IFFT.

It should be emphasized that the present TD--MAS code 
is far from being computationally optimized. We are confident that the application of standard optimization strategies routinely employed in commercial electromagnetic solvers (such as COMSOL or Lumerical), including improved memory management and optimized linear-algebra routines, can significantly reduce both CPU time and RAM usage. Still, even in its current non-optimized form, TD--MAS provides substantially reduced simulation time compared to both FD--COMSOL and FD--MAS, while delivering the complete broadband transient response within a single simulation.

\section{Conclusions}
\label{sec:conclusions}
This work presented a comprehensive time-domain (TD) implementation of the Method of Auxiliary Sources (MAS) combined with the Generalized Sheet Transition Condition (GSTC) for the analysis of two-dimensional planar metasurfaces. The formulation, derived from the impedance-type GSTC, was expressed in convolutional form to ensure causality and temporal accuracy. Through detailed theoretical derivations and a series of numerical experiments, we demonstrated that the proposed TD-MAS--GSTC approach models accurately field interactions across metasurfaces with complex surface responses. Validation was performed using representative scenarios, including magnetostatically biased anisotropic graphene, black \hlg{phosphorus}, and an artificial Lorentzian metasurface. The results confirmed the accuracy, stability, and versatility of the method when benchmarked against frequency-domain MAS and COMSOL simulations. In addition, practical implementation guidelines were provided to facilitate reliable parameter selection and ensure convergence. 

The proposed TD--MAS method is suitable for problems involving electrically thin metasurfaces placed within open or homogeneous environments, where the electromagnetic response is mainly determined by surface interactions. A critical fact is that only the metasurface boundary is discretized, while the fields in the surrounding space are obtained analytically from the radiation of the auxiliary sources. As a result, broadband transient responses can be computed without discretizing the exterior region, which makes the method attractive for problems involving large free-space domains or broadband excitations. On the other hand, for problems involving strongly inhomogeneous media or complex three-dimensional structures, grid-based numerical methods may be more appropriate. The TD--MAS should, therefore, be viewed as an efficient complementary tool for TD analysis of dispersive metasurfaces rather than a replacement of existing numerical techniques.

Future work will focus on extending the TD-MAS--GSTC formulation to spatially dispersive and eventually spacetime-modulated metasurfaces.

\appendices
\section{Discretized TD--MAS--GSTC Matrix Equation}

\setcounter{equation}{0}
\renewcommand{\theequation}{A.\arabic{equation}}

The discretized TD-MAS-GSTC is written in a compact
form via the following abbreviations for the convolution terms

\begin{align*}
\mathcal{A}^{\,i,k'}_{mn,\alpha}
&\triangleq
\sum_{j=0}^{k'} a_{mn}\!\bigl((i-j-k')\Delta t\bigr)\, ZE^{(j)}_{\alpha},\\
\mathcal{B}^{\,i,k'}_{mn,\alpha}
&\triangleq
\sum_{j=0}^{k'} b_{mn}\!\bigl((i-j-k')\Delta t\bigr)\, ZH^{(j)}_{\alpha},\\
\mathcal{C}^{\,i,k'}_{mn,\alpha}
&\triangleq
\sum_{j=0}^{k'} c_{mn}\!\bigl((i-j-k')\Delta t\bigr)\, ZH^{(j)}_{\alpha},\\
\mathcal{D}^{\,i,k'}_{mn,\alpha}
&\triangleq
\sum_{j=0}^{k'} d_{mn}\!\bigl((i-j-k')\Delta t\bigr)\, ZE^{(j)}_{\alpha}.
\end{align*}

\noindent The instantaneous contributions correspond to the special case

\begin{align*}
\mathcal{A}^{0,0}_{mn,\alpha}=a_{mn}(0)ZE^{(0)}_{\alpha},\quad
\mathcal{B}^{0,0}_{mn,\alpha}=b_{mn}(0)ZH^{(0)}_{\alpha},\\
\mathcal{C}^{0,0}_{mn,\alpha}=c_{mn}(0)ZH^{(0)}_{\alpha},\quad
\mathcal{D}^{0,0}_{mn,\alpha}=d_{mn}(0)ZE^{(0)}_{\alpha}.
\end{align*}

\noindent Lastly, for the incident file convolution sums, we have

\begin{align*}
\mathcal{V}^{\,i}_{a_{mn},E}
&\triangleq
\sum_{k'=0}^{i} a_{mn}\!\bigl((i-k')\Delta t\bigr)\, V_E^{(k')},\\
\mathcal{V}^{\,i}_{b_{mn},H}
&\triangleq
\sum_{k'=0}^{i} b_{mn}\!\bigl((i-k')\Delta t\bigr)\, V_H^{(k')},\\
\mathcal{V}^{\,i}_{d_{mn},E}
&\triangleq
\sum_{k'=0}^{i} d_{mn}\!\bigl((i-k')\Delta t\bigr)\, V_E^{(k')},\\
\mathcal{V}^{\,i}_{c_{mn},H}
&\triangleq
\sum_{k'=0}^{i} c_{mn}\!\bigl((i-k')\Delta t\bigr)\, V_H^{(k')}.
\end{align*}

\begin{figure*}[t]
\centering
\tiny 
\begin{multline}
\begin{bmatrix}
\mathcal{A}^{0,0}_{11,1y}+\mathcal{B}^{0,0}_{12,1z}-ZH^{(0)}_{1z}/\Delta t &
\mathcal{A}^{0,0}_{11,2y}+\mathcal{B}^{0,0}_{12,2z}+ZH^{(0)}_{2z}/\Delta t &
\mathcal{A}^{0,0}_{12,1z}+\mathcal{B}^{0,0}_{11,1y} &
\mathcal{A}^{0,0}_{12,2z}+\mathcal{B}^{0,0}_{11,2y}
\\
\mathcal{A}^{0,0}_{21,1y}+\mathcal{B}^{0,0}_{22,1z} &
\mathcal{A}^{0,0}_{21,2y}+\mathcal{B}^{0,0}_{22,2z} &
\mathcal{A}^{0,0}_{22,1z}+\mathcal{B}^{0,0}_{21,1y}+ZH^{(0)}_{1y}/\Delta t &
\mathcal{A}^{0,0}_{22,2z}+\mathcal{B}^{0,0}_{21,2y}-ZH^{(0)}_{2y}/\Delta t
\\
\mathcal{D}^{0,0}_{11,1y}+\mathcal{C}^{0,0}_{12,1z} &
\mathcal{D}^{0,0}_{11,2y}+\mathcal{C}^{0,0}_{12,2z} &
\mathcal{D}^{0,0}_{12,1z}+ZE^{(0)}_{1z}/\Delta t+\mathcal{C}^{0,0}_{11,1y} &
\mathcal{D}^{0,0}_{12,2z}-ZE^{(0)}_{2z}/\Delta t+\mathcal{C}^{0,0}_{11,2y}
\\
\mathcal{D}^{0,0}_{21,1y}-ZE^{(0)}_{1y}/\Delta t+\mathcal{C}^{0,0}_{22,1z} &
\mathcal{D}^{0,0}_{21,2y}+ZE^{(0)}_{2y}/\Delta t+\mathcal{C}^{0,0}_{22,2z} &
\mathcal{D}^{0,0}_{22,1z}+\mathcal{C}^{0,0}_{21,1y} &
\mathcal{D}^{0,0}_{22,2z}+\mathcal{C}^{0,0}_{21,2y}
\end{bmatrix}
\begin{bmatrix}
K_1^{(i)}\\ K_2^{(i)}\\ I_1^{(i)}\\ I_2^{(i)}
\end{bmatrix}
\\
=
-\sum_{k'=0}^{i-1}
\begin{bmatrix}
\mathcal{A}^{i,k'}_{11,1y}+\mathcal{B}^{i,k'}_{12,1z}-ZH^{(i-k')}_{1z}/\Delta t &
\mathcal{A}^{i,k'}_{11,2y}+\mathcal{B}^{i,k'}_{12,2z}+ZH^{(i-k')}_{2z}/\Delta t &
\mathcal{A}^{i,k'}_{12,1z}+\mathcal{B}^{i,k'}_{11,1y} &
\mathcal{A}^{i,k'}_{12,2z}+\mathcal{B}^{i,k'}_{11,2y}
\\
\mathcal{A}^{i,k'}_{21,1y}+\mathcal{B}^{i,k'}_{22,1z} &
\mathcal{A}^{i,k'}_{21,2y}+\mathcal{B}^{i,k'}_{22,2z} &
\mathcal{A}^{i,k'}_{22,1z}+\mathcal{B}^{i,k'}_{21,1y}+ZH^{(i-k')}_{1y}/\Delta t &
\mathcal{A}^{i,k'}_{22,2z}+\mathcal{B}^{i,k'}_{21,2y}-ZH^{(i-k')}_{2y}/\Delta t
\\
\mathcal{D}^{i,k'}_{11,1y}+\mathcal{C}^{i,k'}_{12,1z} &
\mathcal{D}^{i,k'}_{11,2y}+\mathcal{C}^{i,k'}_{12,2z} &
\mathcal{D}^{i,k'}_{12,1z}+ZE^{(i-k')}_{1z}/\Delta t+\mathcal{C}^{i,k'}_{11,1y} &
\mathcal{D}^{i,k'}_{12,2z}-ZE^{(i-k')}_{2z}/\Delta t+\mathcal{C}^{i,k'}_{11,2y}
\\
\mathcal{D}^{i,k'}_{21,1y}-ZE^{(i-k')}_{1y}/\Delta t+\mathcal{C}^{i,k'}_{22,1z} &
\mathcal{D}^{i,k'}_{21,2y}+ZE^{(i-k')}_{2y}/\Delta t+\mathcal{C}^{i,k'}_{22,2z} &
\mathcal{D}^{i,k'}_{22,1z}+\mathcal{C}^{i,k'}_{21,1y} &
\mathcal{D}^{i,k'}_{22,2z}+\mathcal{C}^{i,k'}_{21,2y}
\end{bmatrix}
\begin{bmatrix}
K_1^{(k')}\\ K_2^{(k')}\\ I_1^{(k')}\\ I_2^{(k')}
\end{bmatrix}
\\
+
\begin{bmatrix}
-\mathcal{V}^{\,i}_{a_{12},E}-\mathcal{V}^{\,i}_{b_{11},H}
\\
-\mathcal{V}^{\,i}_{a_{22},E}-\mathcal{V}^{\,i}_{b_{21},H}-V_H^{(i)}/\Delta t
\\
-\mathcal{V}^{\,i}_{d_{12},E}-V_E^{(i)}/\Delta t-\mathcal{V}^{\,i}_{c_{11},H}
\\
-\mathcal{V}^{\,i}_{d_{22},E}-\mathcal{V}^{\,i}_{c_{21},H}
\end{bmatrix}.
\label{eq:matrix}
\end{multline}
\end{figure*}

\noindent The abbreviated discretized TD-MAS-GSTC is given by (\ref{eq:matrix}).


\newpage
\bibliographystyle{IEEEtran}
{\small
\bibliography{mybib_rev_kats}
}

\end{document}